\pdfoutput=1

\documentclass[%
reprint,
 amsmath,amssymb
]{revtex4-2}

\usepackage[pdftex]{graphicx}
\usepackage[hang,small,bf]{caption}
\usepackage[subrefformat=parens]{subcaption}
\usepackage{dcolumn}

\usepackage{bm}

\begin{document}


\title{Systematic study of hadronic excitation energy using Schottky anomaly}

\author{Harunobu Akiyama}
\author{Daisuke Jido}%
\affiliation{%
Department of Physics, Tokyo Institute of Technology
}%

\date{\today}

\begin{abstract}
We extract the excitation energy scales of the hadron spectra in a less model-dependent method using Schottky anomaly. Schottky anomaly is a thermodynamical phenomenon that the specific heat of a system consisting of a finite number of energy levels has a peak at finite temperature due to the energy gaps. Using the masses of all hadrons that are experimentally established, we obtain the excitation energy scales of the hadron spectra and investigate their flavor dependence.   
\end{abstract}

\maketitle

\section{Introduction}
Understanding the hadron structure is one of the important subjects in hadron physics. In particular, the quest of effective constituents of hadrons can be a clue to investigate the hadron structure. For instance, the constituent quark explains the fine structure of the quarkonia \cite{1,2} and the static properties, such as magnetic moment, of the light hadrons \cite{3,4}, even though it is not one of the fundamental elements of quantum chromodynamics. To identify effective degrees of freedom in composite objects, it is good to investigate excitation modes. Recently it has been pointed out in Refs.\ \cite{5,6} that if one takes a diquark picture for the $\Lambda_c$ baryon the diquark-quark confinement potential should be weaker than the quark-antiquark potential obtained in heavy quarkonia in order to reproduce the $\Lambda_c$ excitation spectrum. 
  
  In this work, we visualize the excitation energy scales in various hadronic spectra in a less model-dependent way by examining specific heats obtained by hadronic mass spectra. This method is used widely in different areas.  Reference \cite{7} first applied for hadron physics to extract the effective degrees of freedom of the constituent of a hadron. This is based on the fact that the specific heat of a system is closely related to dynamical degrees of freedom in thermodynamics. This idea was also applied for nuclear systems to investigate their collective degrees of freedom \cite{8}. Recently Ref.\ \cite{schotano} investigated hadronic specific heats calculated by observed heavy meson spectra in order to extract different energy scales underlying the internal dynamics of hadron resonances for the identification of exotic hadrons. By following Ref.\ \cite{schotano}, we extent the study to all of the observed hadrons, and investigate the hadron structure in a model independent way by comparing the visualized excitation energy scales systematically.  In the hadron spectra, excitations in the confinement potential are major modes, while excitations induced by spin dependent forces, such as spin-spin and spin-orbit interactions, are minor and considered as fine structure. In this paper we try to extract such major excitation modes in order to investigate the nature of the confinement force. 
  
The paper is organized as follows.  In Sec.\ II, we explain the characteristics of Schottky anomaly for two-state and $N$-state systems, and apply it for a hadronic model described with the Coulomb-plus-linear potential.  In Sec.\ III, we perform calculations to visualize the excitation energy scales of hadrons using Schottky anomaly, and discuss their flavor dependence.  Finally, Sec.\ IV is devoted to a summary of this work and future prospects. 

\section{Schottky anomaly}

Schottky anomaly is a phenomenon in which the specific heat has a peak called Schottky peak at a finite temperature. The temperature dependence of the specific heat is determined by excitation energies of the system when the number of energy levels in the system is finite.  Now we focus on  low excited states because observed hadrons are in low-lying states.  In this section, we discuss Schottky anomaly using some examples.  First, as one of the simplest examples, we will discuss the two-state system.  Next, we investigate the characteristics of Schottky anomaly using more general systems.  Finally, we will discuss the Schottky anomaly seen in a model close to observed hadronic systems.

\subsection{Two-state system}

\begin{figure}[htbp]
  \begin{center} 
    \includegraphics[width=7.4cm]{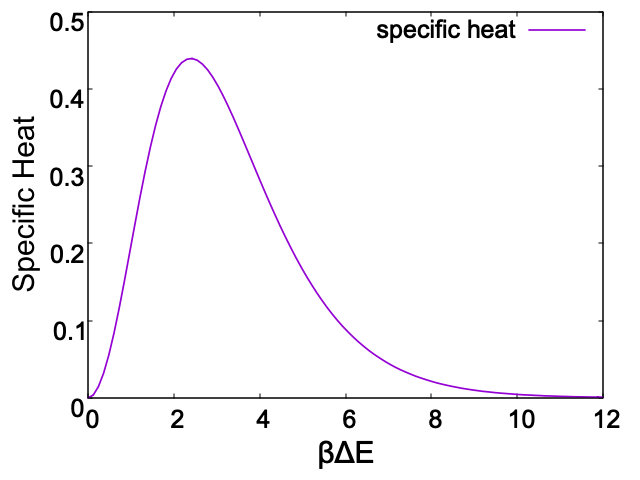}
  \end{center}  
    \caption{Specific heat of the two-state system as a function of $\beta \Delta E$. The Schottky peak is seen at $\beta \Delta E\approx 2.4$.} 
    \label{fig:sh} 
\end{figure}

Let us consider a two-state system with the ground state energy $E_{0}$ and the energy gap $\Delta E$. The canonical partition function of the system is obtained as
\begin{equation}
Z(\beta) = e^{-\beta E_{0}} (1 + e^{-\beta \Delta E}), \label{eq:2state}
\end{equation}
where $\beta$ is the inverse temperature.  The expectation value of the energy of the system can be written as
\begin{align}
\langle E \rangle = -\frac{1}{Z} \frac{\partial Z}{\partial \beta}  \label{eq:E},
\end{align}
using the partition function $Z$.  With this energy expectation value the specific heat of the system is written as
\begin{align}
C &= \frac{\partial }{\partial T} \langle E \rangle \notag \\
&= k_{B} \beta^{2} \frac{\partial^{2} }{\partial \beta^{2}} \ln{Z} \label{eq:C},
\end{align}
for $\beta =1/k_{B} T$.  Hereafter, we take unit of $k_B = 1$, and in this unit the specific heat is a dimensionless quantity, and the temperature $T$ has the dimension of energy.  From Eqs.\ (\ref{eq:2state}) and (\ref{eq:C}), the specific heat of the two-state system is obtained as
\begin{align}
C= (\beta \Delta E)^{2} \left( 2 \cosh{\frac{\beta \Delta E}{2}} \right)^{-2}. \label{eq:2ssh}
\end{align}
As can be seen in Eq.\ (\ref{eq:2ssh}), the specific heat of the two-state system is given as a function of $\beta \Delta E$ and is independent on the energy of the ground state.  The limitting values of the specific heat at sufficiently high and low temperatures are both 0.  Therefore, the specific heat  has a maximum value at a finite temperature.  Figure {\ref{fig:sh}} plots the specific heat $C$ in the two-level system against $\beta \Delta E$.  The specific heat takes a maximum value when $\beta \Delta E \simeq 2.4$ and converges zero at sufficiently high and low temperature.  This behavior is called Schottky anomaly, and this type of the specific heat is called Schottky-type specific heat.  If one observes the peak temperature of the specific heat, $T_{\rm{peak}}$, one can extract the excitation energy by using $\Delta E = 2.4 T_{\rm{peak}}$.

The qualitative interpretation is as follows. For a two-level system with energies $E_0$ and $E_1$, the probability to take the ground state $E=E_0$ at sufficiently low temperature is unity, Prob$(E=E_0)=1$. If the temperature increases, the probability to take the excited state, Prob$(E=E_1)$, becomes large. For the transition from $E_0$ to $E_1$ the system receives heat from the exterior to excite the ground state with the energy gap $\Delta E$, and consequently the specific heat gets large. If the temperature raises further, the probabilities for taking these two states become equal, Prob$(E=E_0)=$Prob$(E=E_1)$, and the system does not need to receive heat from the outside. In this way, the specific heat increases and takes a peak at certain finite temperature corresponding to the energy gap $\Delta E$. 

\subsection{$N$-state systems\label{sec:nstate}}

Let us consider the specific heat of a system with several states and discuss the effect of the multiple energy gaps in the specific heat.  The energy of the $k$th state is given by
\begin{align}
E_{k} = E_{0} + \Delta E_{k} \quad \text{($k=0, \cdots ,N-1$)}, \label{eq:genenergy}
\end{align}
and $\Delta E_{k}$ is the excitation energy of the $k$th state from the ground state and $\Delta E_{0}=0$.

Now, in proceeding with the discussion here, we show that the specific heat of the system generally does not depend on the value of the ground state energy, but only on the excitation energies.  Using Eqs.(\ref{eq:C}) and (\ref{eq:genenergy}) the partition function of the system can be written as
\begin{align}
Z(\beta) &= \sum_{k=0}^{N-1} e^{-\beta E_{k}} \notag \\
&= e^{-\beta E_{0}} \sum_{k=0}^{N-1} e^{-\beta \Delta E_{k}},
\end{align}
From Eq.\ (\ref{eq:E}), the expected value of energy is
\begin{align}
\langle E \rangle = E_{0} - \frac{1}{z} \frac{\partial z}{\partial \beta} \label{eq:gs},
\end{align}
where a reduced partition function $z$ is defined by
\begin{align}
z(\beta) &\equiv \sum_{k=0}^{N-1} e^{-\beta \Delta E_{k}}.
\end{align}
The specific heat is obtained by differentiating both sides of Eq.\ (\ref{eq:gs}) with temperature.  At this time, the first term of the right hand side of Eq.\ (\ref{eq:gs}) does not contribute to the specific heat because it is a constant. The second term does not depend on the ground state energy $E_{0}$.  Therefore, the specific heat of the system is determined only by the excitation energies, and independent of the ground state energy. 

\begin{figure}[htbp]
\begin{minipage}[b]{\linewidth}
\centering
\includegraphics[width=7.4cm]{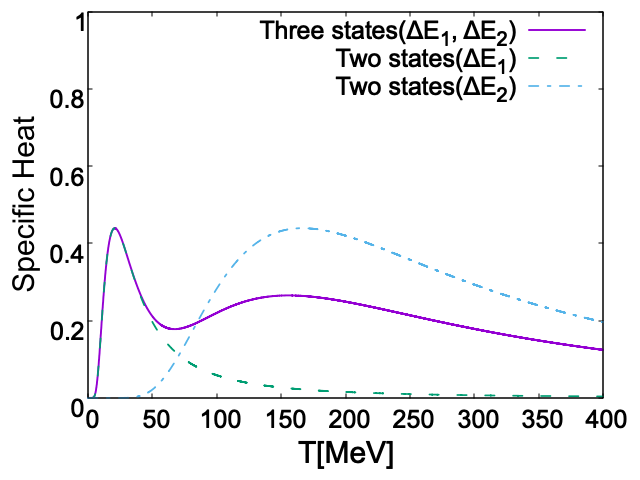}
\subcaption{$\Delta E_{1} = 50$MeV \label{50400}}
\end{minipage}\\
\begin{minipage}[b]{\linewidth}
\centering
\includegraphics[width=7.4cm]{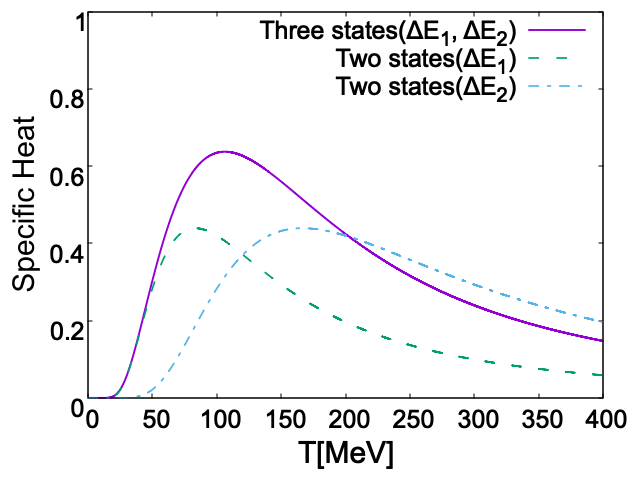}
\subcaption{$\Delta E_{1} = 200$MeV \label{200400}}
\end{minipage}\\
\caption{The solid line in each figure shows specific heat when there are two excitation energies.  Dashed lines and dashed-dotted line shows the specific heat when $\Delta E = \Delta E_{1}, \Delta E_{2}$(=400MeV) with one excitation energy.}
\label{fig:mscale}
\end{figure}

  We first consider the case for $N=3$, in which there are two excitation energies, $\Delta E_1$ and $\Delta E_2$. In this system, the reduced partition function $z$ is given as
\begin{align}
z(\beta) = 1 + e^{-\beta \Delta E_{1}} + e^{-\beta \Delta E_{2}}.
\end{align}
  Let us fix the second excitation energy as $\Delta E_2 = 400$~MeV. We compere the specific heats calculated with $\Delta E_1 = 50$~MeV and $\Delta E_1 = 200$~MeV. In the former case, two excitation energies have different magnitudes, while in the latter they are comparable. In Fig.~\ref{fig:mscale}, we plot the specific heats for $\Delta E_1 = 50$~MeV (a) and $\Delta E_1 = 200$~MeV (b). The solid lines show the specific heats of the three-state systems. For comparison, we also plot the specific heats of the two-state systems with $\Delta E_1$ and $\Delta E_2$ as the dashed and dashed-dotted lines, respectively. In Fig.~\ref{50400} there are two separated peaks. Comparing it with the specific heat for the two-state systems, one finds that the peak at lower temperature corresponds to the smaller excitation energy $\Delta E_1$, while the peak at higher temperature to the larger excitation energy $\Delta E_2$. It is also seen in the figure that the height of the peak at the lower temperature is higher than that of the other. This implies that the specific heat of the three-state system is dominated by the lower excitation mode. Figure~\ref{200400} shows the specific heat of the three-state system with $\Delta E_1 = 200$~MeV. Here one finds only one peak. It is interesting to note that the peak position is just between the temperatures corresponding to the excitation energies $\Delta E_1$ and $\Delta E_2$. Thus, when two excitation energies have similar values, which means that two level have a similar energy, the specific heat has one peak at a temperature corresponding to an energy between two excitation energies. In this way we can extract an isolated excitation energy scale using Schottky anomaly without looking at fine energy gaps. 

\begin{figure}[htbp]
  \begin{center} 
    \includegraphics[width=7.4cm]{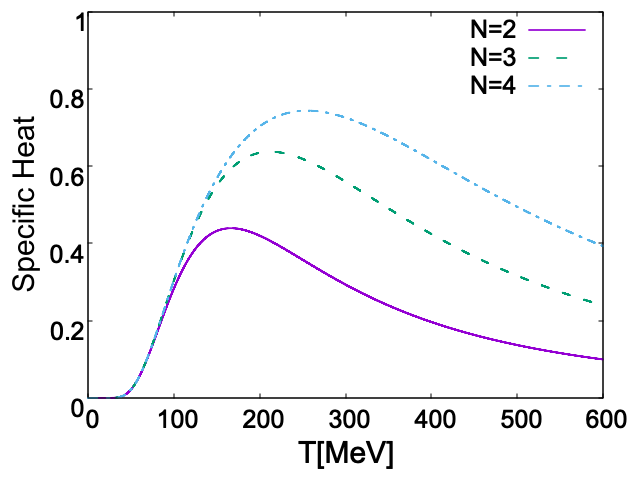}
    \caption{Specific heats of the $N$-state systems with an equal energy gap $\Delta E= 400$ MeV. The solid, dotted and dashed-dotted lines denote the specific heats of the 2-state, 3-state and 4-state systems, respectively.} 
    \label{fig:trc1} 
  \end{center}
\end{figure}

Next we demonstrate the effect on the Schottky peak by higher excited states. For this purpose, we consider $N$-state systems with an equal energy gap, $\Delta E_k = k \Delta E$ with $\Delta E = 400$~MeV. Figure~\ref{fig:trc1} shows the specific heats for a 2-state system (solid line), a 3-state system (dashed line) and a 4-state system (dotted-dashed line). The peak temperatures can be read as 167~MeV, 213~MeV and 256~MeV for the 2-state, 3-state and 4-state systems, respectively. The excitation energy scale of these systems is 400~MeV. This can be extracted in the 2-state system by reading the temperature of the peak position and multiplying it by 2.4. In the 3-state system, the Schottky peaks by these two excitation modes are overlapped and only one peak appears and the peak position is shifted about 20\% higher. For the 4-state system, similarly, the peak position is shifted about 20\% higher than that of the 3-state system. In this way, if there is a series of excitation modes, the effect of higher excited states on the peak position is seen in a 20\% shift of the peak position. This tells us that the estimate of the energy scale by Schottky anomaly has potentially about 20\% error. 

It is interesting to point out that the specific heat of the $N$-state systems with an equal energy gap can be expressed as a function of dimensionless parameter $x=\beta \Delta E$ as
\begin{align}
C &= x^{2} \frac{\sum_{k}\sum_{l}k^{2}e^{-(k+l)x } - \left( \sum_{k} ke^{-kx} \right)^{2}}{\left( \sum_{k} e^{-kx} \right)^{2}}. 
\end{align}
Therefore, the peak positions of the specific heat are determined by $x$ independently of the explicit value of $\Delta E$. Because the temperature is scaled as $T = x \Delta E$, the estimation error discussed above is independent of the explicit value of $\Delta E$.

\subsection{\label{sec:calccbar}Analysis of calculated data}

\begin{figure}[htbp]
\begin{minipage}[b]{\linewidth}
\centering
\includegraphics[width=7.4cm]{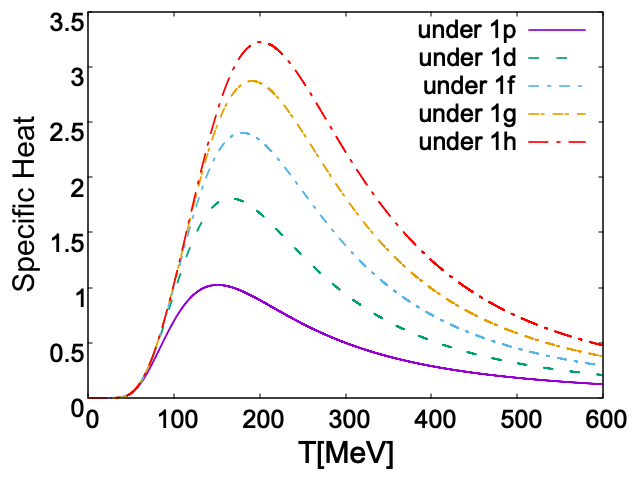}
\subcaption{$k=0.9\rm{GeV\cdot fm^{-1}}$}
\end{minipage}\\
\begin{minipage}[b]{\linewidth}
\centering
\includegraphics[width=7.4cm]{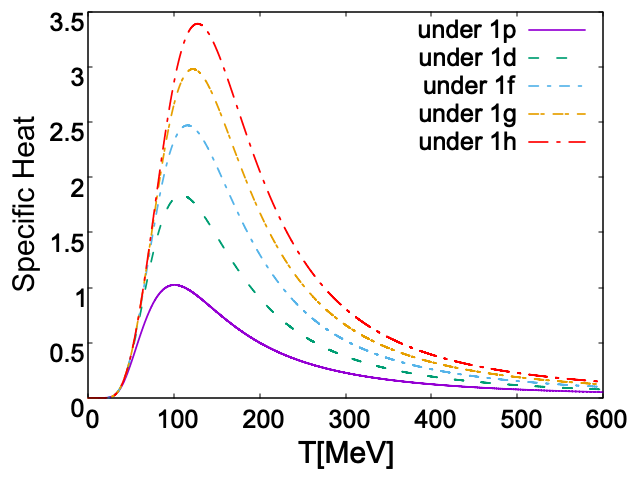}
\subcaption{$k=0.4\rm{GeV\cdot fm^{-1}}$}
\end{minipage} 
\caption{Specific heat of the charmonium system calculated by a quark model with potential (\ref{eq:potential}).\label{fig:trc2} }
\end{figure}

In this subsection, we consider more a realistic spectrum using a potential model for the heavy quarkonia and investigate the behavior of the specific heat.  The charmonium spectrum can be reproduced well by a central potential given by 
\begin{align}
V(r) = -\frac{4}{3}\frac{\alpha_{s}}{r} + kr + V_{0}, \label{eq:potential}
\end{align}
with spin dependent forces as long as the $\bar DD$ threshold does not open \cite{2}. Here, $\alpha$, $k$ and $V_0$ are the potential parameters to be fitted by experimental observables.  Here we use a parametrization given in Ref.~\cite{5} as an example. We calculate the energy eigenvalue of charmonia using only the central potential. The eigenvalue can be obtained by solving the eigenvalue equation
\begin{equation}
-\frac{1}{2\mu}\frac{d^{2}\chi_{\ell}(r)}{dr^{2}} + \left[ V(r) + \frac{\ell(\ell + 1)}{2\mu r^{2}} \right]\chi_{\ell}(r) = E_{n,\ell} \chi_{\ell}(r),
\end{equation}
where $\mu$ is the reduced mass of the charm and anti-charm quarks. The charm quark mass is set as 1.5 GeV.  Table~\ref{tab:calc} shows the energy eigenvalues obtained in the calculation.  The explicit values of the parameters that we use are $k = 0.9$~$\rm{GeV\cdot fm^{-1}}$ and $\alpha=0.4$. Parameter $V_0$ is adjusted so that the ground state energy is to be zero. We also calculate the charmonium spectrum using $k = 0.4$~$\rm{GeV\cdot fm^{-1}}$ in order to check whether the Schottky peak can distinguish the strength of the string tension.  The partition function of this system can be written as
\begin{align}
Z(\beta) = \sum_{n,\ell}(2\ell +1)e^{-\beta E_{n.\ell}}.
\end{align}
Here, the coefficient is the degeneracy due to the magnetic quantum number.  The specific heats calculated from these spectra are shown in Fig.~\ref{fig:trc2}.  We summarize the values of the peak temperature $T_{\rm{peak}}$ in Table~\ref{tab:tpeakcalc}.  Comparing the results for different values of $k$, one can see that the Schottky peak can distinguish the size of the potential parameter.

In Sec.~III, we will calculate the specific heats of the observed hadron spectra to obtain the excitation energies. There we consider the observed hadron spectra, which have up to the first or second excited states. We obtain the excitation energy scale by using $\Delta E = 2.4T_{\rm{peak}}$ following the two-state system. Here, we verify the validity of this argument by performing a similar calculation using the spectrum obtained by the model calculation with $k=0.9$~$\rm{GeV\cdot fm^{-1}}$. First, we obtain $\Delta E = 362$~MeV as a result of calculating the excitation energy scale using the states up to $1P$. This is about 20\% different from 430~MeV that is the energy gap of in the ground state and the $1P$ state. This difference comes from the degeneracy of the 1P states due to the orbital angular momentum $\ell=1$. In the application for the observed spectra, we extract the excitation energies without specifying the angular momentum of the excited states by using $\Delta E = 2.4T_{\rm{peak}}$. Thus this may underestimate the excitation energy about 20\% and we accept this difference as uncertainty of the estimation. We also discuss the extraction uncertainty coming from the number of the observed states. Let us take the potential (\ref{eq:potential}) again as an example. If we consider the excited states up to the $1D$ state of the potential (\ref{eq:potential}) , the peak temperature for this spectrum is about 10\% larger than that for the spectrum up to the $1P$ state according to Table~\ref{tab:tpeakcalc}. Similarly the peak temperature for the spectrum up to the $1F$ state is about 10\% larger than that up to $1D$ state. Considering the fact that only some lower excited states have been observed, we evaluate the uncertainty of the estimation of the excitation energies to be 20\%.

\begin{table}
\caption{Calculated energy eigenvalue of charmonium for $k=0.9,\ 0.4$ $\rm{GeV\cdot fm^{-1}}$ in units of MeV, using $\alpha_{s} = 0.4$, $\mu = 0.75$ $\rm{GeV}$. \label{tab:calc}}
\begin{ruledtabular}
\begin{tabular}{lll}
state & $k=0.9$ $\rm{GeV\cdot fm^{-1}}$ & $k=0.4$ $\rm{GeV\cdot fm^{-1}}$  \\ \hline 
$1S$ & 0 & 0 \\
$1P$ & 430 & 287 \\
$2S$ & 604 & 384 \\
$1D$ & 723 & 468 \\
$2P$ & 890 & 564 \\
$1F$ & 970 & 617 \\
$3S$ & 1043 & 648 \\
$2D$ & 1125 & 707 \\
$1G$ & 1191 & 709 \\
$3P$ & 1276 & 793 \\
$2F$ & 1335 & 833 \\
$1H$ & 1395 & 870 \\
\end{tabular}
\end{ruledtabular}
\end{table}

\begin{table}
\caption{Peak temperature of the specific heats in units of MeV.\label{tab:tpeakcalc}}
\begin{ruledtabular}
\begin{tabular}{lllll}
&\multicolumn{2}{l}{$k=0.9\ \rm{GeV\cdot fm^{-1}}$}&\multicolumn{2}{l}{$k=0.4\ \rm{GeV\cdot fm^{-1}}$} \\
States & $T_{\rm{peak}}$ & 2.4$T_{\rm{peak}}$  & $T_{\rm{peak}}$ & 2.4$T_{\rm{peak}}$ \\ \hline 
Up to $1P$ & 151 & 362 & 101 & 242\\
Up to $1D$ & 166 & 398 & 109 & 262\\
Up to $1F$ & 180 & 432 & 116 & 278\\
Up to $1G$ & 191 & 458 & 122 & 293\\
Up to $1H$ & 202 & 485 & 128 & 307\\
\end{tabular}
\end{ruledtabular}
\end{table}

\section{Application to hadron}
In this chapter, we apply the Schottky anomaly for observed hadron spectra and visualize a typical hadron excitation energy scale for each hadron spectrum.  Furthermore, we investigate the hadron structure systematically by comparing the excitation energy scales of each spectrum.  We note that the specific heat and temperature treated here do not represent the thermodynamic feature of hadrons but are merely mathematical parameters.  For a hadronic system, the canonical partition function of the system are given as
\begin{align}
Z_{\rm{hadron}}(\beta) = \sum_{i} (2J_{i} + 1)(2I_{i} + 1)e^{-\beta E_{i}},
\end{align}
using observed hadron mass $E_{i}$ with the degeneracy of the total angular momentum $J$ and the isospin $I$.  Here we use the hadron masses without their decay width.  The specific heat of the system is expressed with the partition function as
\begin{align}
C_{\rm{hadron}} = \beta^{2} \frac{\partial^{2}}{\partial \beta^{2}} \ln{Z_{\rm{hadron}}}.
\end{align}
By plotting $C_{\rm{hadron}}$ against temperature we read the Schottky peak temperature $T_{\rm{peak}}$ and the extract excited energy scales of the system by
\begin{align}
\Delta E = 2.4T_{\rm{peak}}.
\end{align} 
Here we presume that the observed hadron spectra have up to first or second excited states.  It is important that the first excited states should be observed enough to obtain reliable results.

For the calculation of the specific heats we use the hadron masses shown by particle data group \cite{pdg} and pick up the hadrons of which total angular moment and isospin have been determined and of which existence is sufficiently confirmed experimentally with $\bullet$ for mesons and **** or *** for baryons in the summary tables. We have summarized the list of states used in the calculations in this chapter in Appendix~\ref{sec:list}.

\subsection{Heavy mesons}
In this section, we calculate the excitation energy scales of heavy mesons using Schottky anomaly.  First we perform calculations for heavy qurkonia and then for open charm and open bottom states.  Finally we discuss the flavor dependence of heavy meson energy scales from these results.  Reference~\cite{schotano} calculated the specific heat of the heavy mesons to search for exotic mesons.

\subsubsection{Heavy qurakonia}
\begin{figure}
\includegraphics[width=7.4cm]{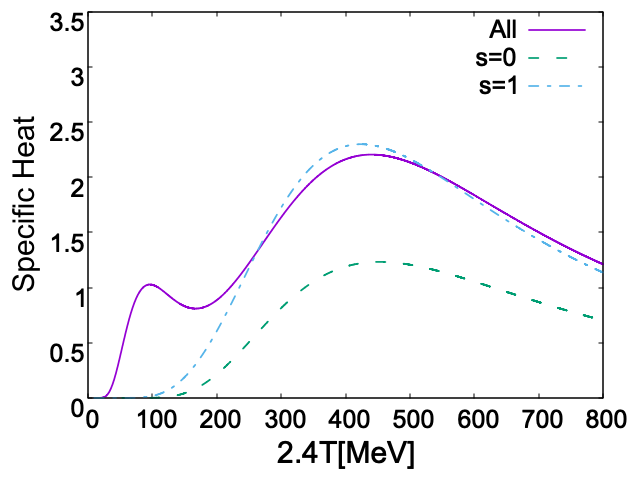}
\caption{Specific heat of charmonium. The solid line show the specific heat of all charmonium system. The dashed and dashed-dotted lines denote the specific heats calculated by using charmonia with $s=0$ and $s=1$ separately, respectively. The separation is done in a way written in the text.  \label{fig:ccbar}}
\end{figure}

In the calculation we use the charmonia listed in Table~\ref{tab:ccbar}. We calculate the specific heat using these charmonium states with the spin degeneracy. Figure~\ref{fig:ccbar} shows the results of calculating the specific heat of the charmonia as a function of $2.4 T$.  The solid line in Fig.~\ref{fig:ccbar} shows the specific heat calculated with all of the charmonia in the list. We find that the solid line has two Schottky peaks. From these peaks we read the excitation energies $\Delta E =$ 106~MeV and 441~MeV for the lower and higher peaks, respectively.  As will be described later, we consider that the peak on the low temperature corresponds to the hyperfine structure due to the spin-spin interaction and the peak on the high temperature corresponds to the orbital excitation due to the central potential.

In order to understand the nature of the excitation energies, we separately plot the specific heats for the charmonia by the spin configuration of the quarks, $s=0$ and $s=1$.  If the charmonia are assumed to be $c\bar{c}$ states, the spin configuration can be specified by parity $P$ and charge conjugation $C$ as follows.  The parity of the meson can be written as 
\begin{align}
P=(-1)^{\ell+1}, \label{eq:parity}
\end{align}
using the orbital angular momentum $\ell$.  This is because orbital motion gives parity $(-1)^\ell$ and $q\bar q$ has negative internal parity in addition.  Next for $C$ parity, because taking charge-conjugation of $q\bar{q}$ is same as the replacement of quark and antiquark, the charge conjugate $C$ is found as
\begin{align}
C=(-1)(-1)^{\ell}(-1)^{s+1}, \label{eq:cparity}
\end{align}
for the orbital angular momentum $\ell$ and the spin $s$. With Eqs.(\ref{eq:parity}) and (\ref{eq:cparity}) one finds
\begin{align}
C=P(-1)^{s+1}. 
\end{align}
This implies that $P=C$ for $s=1$ and $P=-C$ for $s=0$. 

In Fig.~\ref{fig:ccbar} the results of the specific heats for $s=0$ and $s=1$ are shown as short dash line and dash-dotted line, respectively. We see that the lower peak disappears in both plots and find the excitation energies $\Delta E=453$~MeV for $s=0$ and 424~MeV for $s=1$.  This means that the lower peak seen in the full spectrum should correspond to the hyperfine splitting induced by the spin-spin interaction between quark and antiquark, and that the higher peak in the full spectrum is common for both spin configuration.  

\begin{figure}
\includegraphics[width=7.4cm]{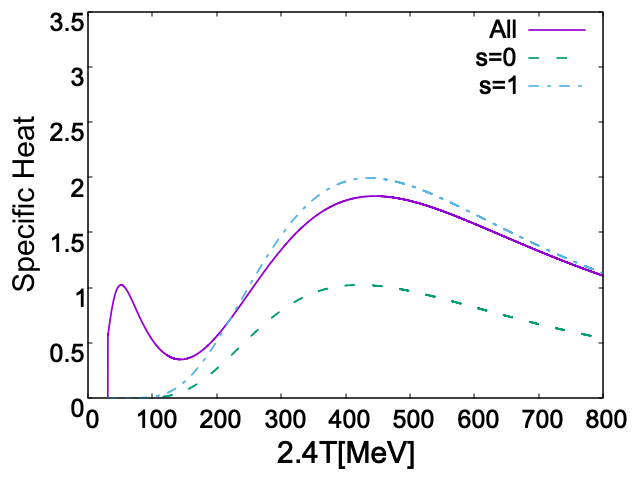}
\caption{Same as in Fig.\ \ref{fig:ccbar} but for bottomonium.\label{fig:bbbar}}
\end{figure}

Let us move to the calculation for bottomnium.  Figure~\ref{fig:bbbar} shows the results of the specific heat for bottomonium. In the calculation we use the bottomonia shown in Table~\ref{tab:bbbar}. The solid line in Fig.~\ref{fig:bbbar} shows the specific heat calculated with all of the bottomonia in the list.  Again, we obtain two excitation energy scales, and the values of each energy scale are 51~MeV and 445~MeV.  Similarly to charmonium, we plot separately the specific heats for $S=0$ and $S=1$ in Fig.~\ref{fig:bbbar} as dashed and dot-dashed lines, respectively.  It is considered that the peak on the low temperature corresponds to the hyperfine structure and the peak on the high temperature corresponds to the orbital excitation, as in the case of charmonium.  The energy scales of orbital excitation of bottomonium is close to 440~MeV like the charmonium.  Therefore, the energy scales of the orbital excitation are insensitive to the flavor for the heavy quarkonia.  On the other hand, the energy scale of the hyperfine structure of bottomonium is about half that of charmonium.  This can be explained by the fact that the magnitude of splitting due to spin-spin interaction can be written as,
\begin{align}
\Delta E_{\rm{hf}} = \frac{32\pi}{9} \frac{\alpha_{s}}{m_{q}m_{\overline{q}}}|\Psi(0)|^{2},  \label{eq:HF}
\end{align}
using quark model \cite{particlephys}, and the splitting decreases as the quark mass increases.  The excitation energy scales obtained here are summarized in Table~\ref{tab:sumhm}.

\subsubsection{Open charm and open bottom mesons}
Here we consider open charm and open bottom mesons.  In the calculation we use the charmed mesons and bottomed mesons shown in Table~\ref{tab:openc} and Table~\ref{tab:openb}, respectively. 

\begin{figure}[htbp]
\begin{minipage}[b]{\linewidth}
\centering
\includegraphics[width=7.4cm]{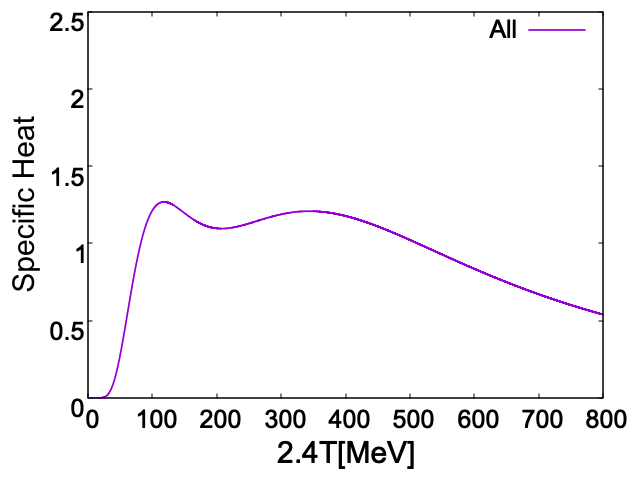}
\subcaption{\label{fig:oc}}
\end{minipage}\\
\begin{minipage}[b]{\linewidth}
\centering
\includegraphics[width=7.4cm]{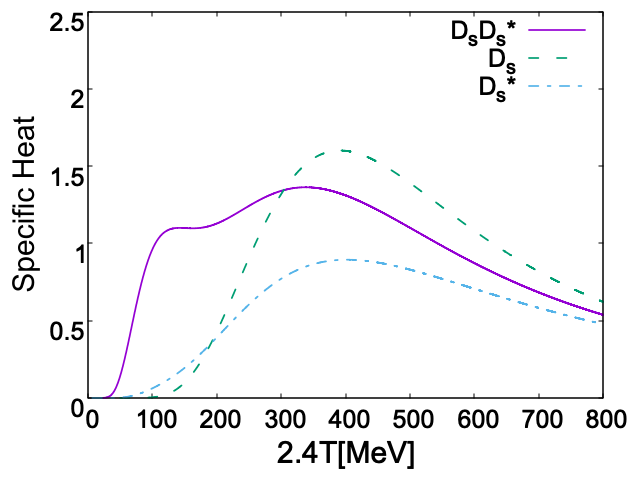}
\subcaption{\label{fig:ocs}}
\end{minipage}\\
\begin{minipage}[b]{\linewidth}
\centering
\includegraphics[width=7.4cm]{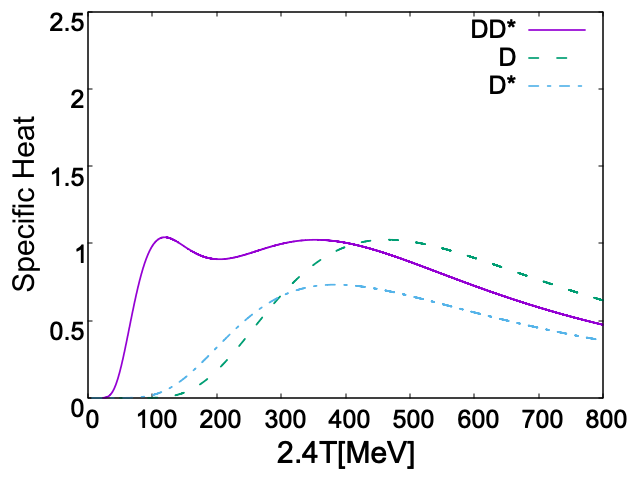}
\subcaption{\label{fig:ocud}}
\end{minipage} \\ 
\caption{Specific heats of the open charm systems. (a) Specific heat calculated with all the open charm states given Table IX. (b) Specific heats of the open charm states with strangeness. The solid line stands for the specific heat of all the strange charm mesons, while the dashed and dashed-dotted lines show the specific heats of the $D_s$ and $D_s^*$ mesons, respectively. (c) Specific heats of the nonstrange open charm states. The solid line stands for the specific heat of all the nonstrange charm mesons, while the dashed and dashed-dotted lines show the specific heats of the $D$ and $D^*$ mesons, respectively.}
\label{fig:openc}
\end{figure}

First, we calculate the specific heat of the open charm system.  Figure~\ref{fig:oc} shows the specific heat calculated with all of the open charm mesons.  There are two peaks again and the corresponding excitation energies are read as 108~MeV and 344~MeV for lower and higher peaks, respectively.  We will see that the low temperature peak is due to the mass difference between the $s$ quark and the $u$ or $d$ quark, or hyperfine structure.  

We plot separately the $D_s$ and $D$ mesons in Fig.~\ref{fig:ocs} and Fig.~\ref{fig:ocud} respectively.  The solid lines in both figures show  the specific heat calculated with all of the $D_s$ and $D$ mesons.  Similar to the charmonium, two peaks appear for both cases, and the values of the energy scale are 119~MeV and 353~MeV for strange open charm and 143~MeV and 338~MeV for nonstrange open charm.  

In order to see the origins of these peaks, we separately calculate the specific heats of the $D_s$ and $D_s^*$ mesons and the $D$ and $D^*$ mesons, in which the $D_s$ and $D$ states have $J^p = 0^-, 1^+, 2^-, \dots$, while the $D_s^*$ and $D^*$ have $J^p = 0^+, 1^-, 2^+, \dots$. Although this is not a classification of the quark spin configuration, we can resolve the splitting of the ground states induced by the spin-spin interaction. The separated specific heats are plotted in Fig.~\ref{fig:ocs} and Fig.~\ref{fig:ocud} as dashed lines for $D_s$ and $D$ and dashed-dotted lines for $D_s^*$ and $D^*$. The figures show only one peak appears at higher temperature for each line. This implies that the peaks appearing at lower temperature in the solid lines of Fig.~\ref{fig:ocs} and Fig.~\ref{fig:ocud} are considered to be the energy scale of the hyperfine splitting as seen in quarkonia. It is notable that the energy scales of the hyperfine splitting read 143~MeV for $D_s$ and 119~MeV for $D$. This is opposite to the quark mass dependence of the hyperfine splitting Eq. (\ref{eq:HF}). This would implies that the overlap of the wavefunction $|\Psi(0)|^2$ for $D_s$ be larger than that for $D$.  Regarding the peaks at higher temperature, the energy scales read 391 MeV for $D_s$, 402~MeV for $D_s^*$, 464~MeV for $D$ and 384~MeV for $D^*$ from the dashed and dashed-dotted lines in Fig.~\ref{fig:ocs} and Fig.~\ref{fig:ocud}. These are larger than the energy scales seen at higher temperature in the solid lines. This means that there could be other energy scales than the orbital excitation in the $D$-$D^*$ and $D_s$-$D_s^*$ spectra.

\begin{figure}[htbp]
\begin{minipage}[b]{\linewidth}
\centering
\includegraphics[width=7.4cm]{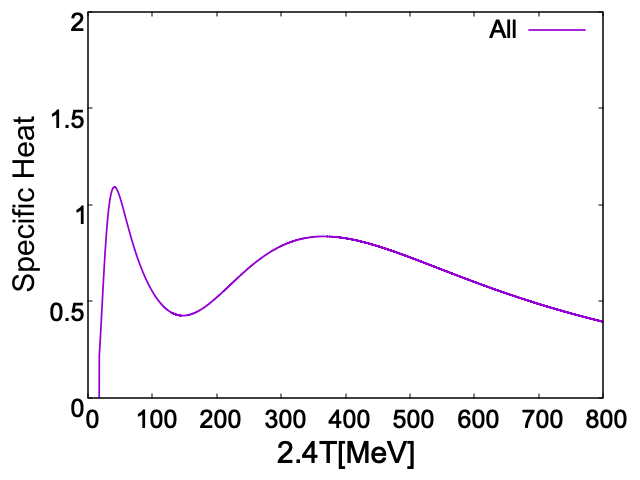}
\subcaption{\label{fig:ob}}
\end{minipage}\\
\begin{minipage}[b]{\linewidth}
\centering
\includegraphics[width=7.4cm]{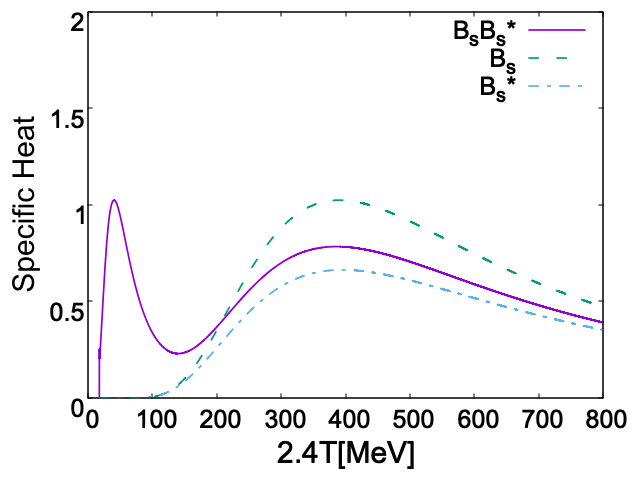}
\subcaption{\label{fig:obs}}
\end{minipage}\\
\begin{minipage}[b]{\linewidth}
\centering
\includegraphics[width=7.4cm]{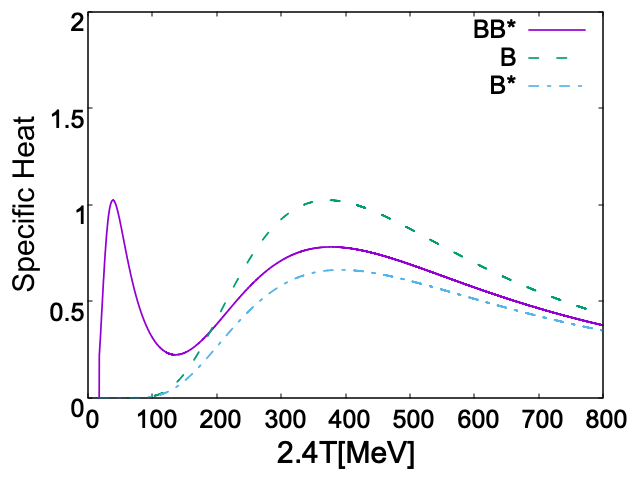}
\subcaption{\label{fig:obud}}
\end{minipage} \\ 
\caption{Same as in Fig.\ \ref{fig:openc} but for the open bottom systems.}
\label{fig:openb}
\end{figure}

Next, we calculate the specific heat of the open bottom system.  Figure~\ref{fig:ob} shows the specific heat calculated with all of the open bottom mesons and there are two peaks again and the corresponding excitation energies are read as 41~MeV and 367~MeV for lower and higher peaks, respectively.

We plot separately the $B_s$ and $B$ mesons in Fig.~\ref{fig:obs} and Fig.~\ref{fig:obud} respectively and the solid lines in both figures show the specific heat calculated with all of the $B_s$ and $B$ mesons.  Similar to open charm, two peaks appear for both case, and the energy scale values are 40~MeV and 386~MeV for strange open bottom and 39~MeV and 378~MeV for nonstrange open bottom.    

In order to see the origins of these peaks, we separately calculate the specific heats of the $B_s$ and $B_s^*$ mesons and the $B$ and $B^*$ mesons.  The separated specific heats are plotted in Fig.~\ref{fig:obs} and Fig.~\ref{fig:obud} as dashed lines for $B_s$ and $B$ and dashed-dotted lines for $B_s^*$ and $B^*$. The figures show only one peak appears at higher temperature for each line. This implies that the peaks appearing at lower temperature in the solid lines of Fig.~\ref{fig:obs} and Fig.~\ref{fig:obud} are considered to be the energy scale of the hyperfine splitting as seen in open charms.  Regarding the peaks at higher temperature, the energy scales read 390~MeV for $B_s$, 395~MeV for $B_s^*$, 378~MeV for $B$ and 393~MeV for $B^*$ from the dashed and dashed-dotted lines in Fig.~\ref{fig:obs} and Fig.~\ref{fig:obud}. These are larger than the energy scales seen at higher temperature in the solid lines. This means that there could be other energy scales than the orbital excitation in the $B$-$B^*$ and $B_s$ and $B_s^*$ spectra.

\subsubsection{Summary of heavy mesons}
We summerize the excitation energy scale of charmonium, bottomonium, open charm and open bottom system in Table~\ref{tab:sumhm}.  In the heavy meson systems, we have found two energy scales for hyperfine structure by spin-spin interaction and the orbital excitation by central potential.  Comparing the results of charmonium and bottomonium, we find that the energy scale of the orbital excitation is about 440~MeV, and it is insensitive to the flavor of the heavy mesons.  On the other hand, the energy scale of hyperfine structure becomes smaller as the quark mass increases.  For the open charm and bottom systems we obtain a similar result of the quarkonia, that is, the two peaks are found in the specific heats for all the open heavy mesons; the higher peaks correspond to the orbital excitations, while the lower peaks to the hyperfine splitting. After separating the open heavy mesons in terms of strangeness and spin, we find that the excitation energies corresponding to the orbital motion are about 400~MeV independently of their flavor.

\begin{table}
\caption{Excitation energy scales of the heavy mesons obtained by the Schottky anomaly in units of MeV. [MeV] \label{tab:sumhm}}
\begin{ruledtabular}
\begin{tabular}{lllllll}
    & \multicolumn{2}{l}{$c\bar{c}$} & \multicolumn{2}{l}{$C=1,S=1$}  & \multicolumn{2}{l}{$C=1,S=0$}  \\
all & \multicolumn{2}{l}{106, 441}    & \multicolumn{2}{l}{143, 338}    & \multicolumn{2}{l}{119, 353}    \\
    & $s=0$          & $s=1$         & $D_{s}$      & $D_{s}^{*}$     & $D$          & $D^{*}$         \\
    & 453            & 424           & 391          & 404             & 384          & 412             \\ \hline
    & \multicolumn{2}{l}{$b\bar{b}$} & \multicolumn{2}{l}{$B'=1,S=1$} & \multicolumn{2}{l}{$B'=1,S=0$} \\
all & \multicolumn{2}{l}{51, 445}     & \multicolumn{2}{l}{40, 386}     & \multicolumn{2}{l}{39, 378}     \\
    & $s=0$          & $s=1$         & $B_{s}$      & $B_{s}^{*}$     & $B$          & $B^{*}$         \\
    & 446            & 437           & 390          & 395             & 378          & 393            
\end{tabular}\end{ruledtabular}
\end{table}

\subsection{Light mesons}
\begin{figure}[htbp]
\begin{minipage}[b]{\linewidth}
\centering
\includegraphics[width=7.4cm]{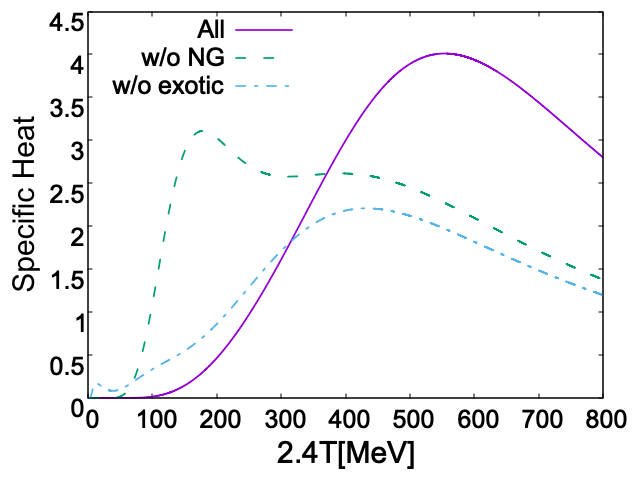}
\subcaption{\label{fig:lm}}
\end{minipage}\\
\begin{minipage}[b]{\linewidth}
\centering
\includegraphics[width=7.4cm]{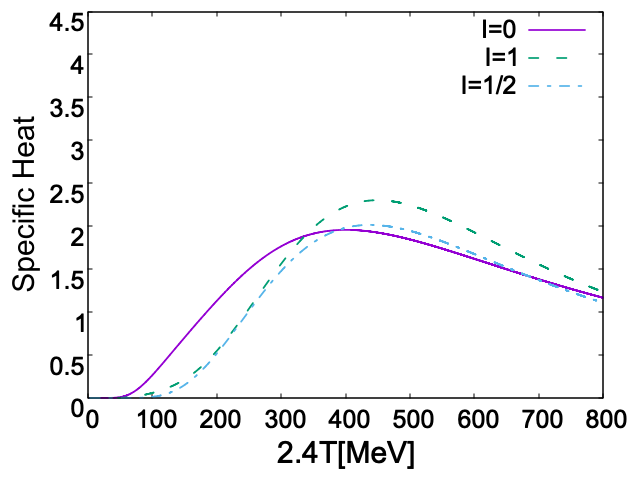}
\subcaption{\label{fig:lmiso}}
\end{minipage} \\ 
\caption{(a) Specific heats for the light mesons. The solid line denotes the specific heat calculated with all the light mesons given in Table XI. The dashed line stands for the specific heat calculated without the Nambu-Goldstone bosons, $\pi$, $K$, $\eta$, while the dashed-dotted line shows the one without the Nambu-Goldstone bosons and the scalar exotics, $f_0(500)$ and $K_0^*(700)$. (b) Specific heats of the light mesons for each isospin. The specific heats are obtained without the Nambu-Goldstone bosons and the scalar exotics. The solid, dashed and dashed-dotted lines show the specific heats of the light mesons with $I=0$, $I=1$ and $I=1/2$, respectively. }
\label{fig:lightmeson}
\end{figure}

In this section, we calculate the specific heat for light mesons.  First of all, let us start with the calculation of the specific heat for all the light mesons summarized in Table~\ref{tab:lightmeson}. For the light mesons, there are several states out of regularity such as Nambu-Goldstone bosons and exotic candidates. These irregular mesons could spoil the systematic extraction of the excitation energies. In Fig.~\ref{fig:lm} we show the specific heat obtained by all the light mesons in the solid line. The plot shows that there is only one peak structure at 554~MeV. This energy scale is much larger than the global excitation energy seen in the heavy mesons. We expect that the Schottky peak in Fig.~\ref{fig:lm} contains several excitation energies. To resolve these energies, we first subtract the Nambu-Goldstone bosons ($\pi$, $K$ and $\eta$) from the spectrum and calculate the specific heat without them. As seen in the dashed line of Fig.~\ref{fig:lm}, there are two peaks at 178~MeV and 393~MeV. Next we remove the exotic candidates, $f_0(500)$ and $K_0^*(700)$, from the calculation. The specific heat is shown in the dashed-dotted line of Fig.~\ref{fig:lm}. This plot shows two excitation energies at 13~MeV and 432~MeV. The lower peak comes from the energy difference between $\rho(770)$ and $\omega(782)$. The position of the higher peak is closed to that of the dashed line. Thus, we regard that the higher excitation energy is a typical excitation energy scale of the light mesons. It is interesting that this energy scale is very similar with what we find in heavy mesons. 

Next we decompose the light mesons in terms of their isospin. Again we remove the Nambu-Goldstone bosons and the two scalar mesons from the calculation. The solid line in Fig.~\ref{fig:lmiso} shows the specific heat of the $I=0$ states and the peak position is read as 401~MeV. Similarly the specific heats of the $I=1$ and $I=1/2$ are shown in the dashed and dashed-dotted lines of Fig.~\ref{fig:lmiso} respectively, and the excitation energies are read as 450~MeV and 438~MeV. These excitation energy scales are similar to that we have obtained using the light mesons before the isospin decomposition. From these results excluding the Nambu-Goldstone bosons and the scalar mesons, the typical excitation energy scale of the light mesons is estimated to be about 400 to 450~MeV. Considering this result together with those of the heavy mesons (Table~\ref{tab:sumhm}), we find flavor independence in the excitation energy scale of the meson systems.

\begin{table}
\caption{Excitation energies of the light mesons obtained by Schottky anomaly in units of MeV.\label{tab:sumlightmeson}}
\begin{ruledtabular}
\begin{tabular}{llll}
                & All  & w/o $\pi, K, \eta$ & w/o $\pi, K, \eta, f_{0}(500), K_{0}^{*}(700)$ \\
All             & 554  & 178, 393          & 13, 432                           \\
$I=0$           &  -   &   -               & 401                               \\
$I=1$           &  -   &   -               & 450                               \\
$I=1/2$         &  -   &   -               & 438                              
\end{tabular}
\end{ruledtabular}
\end{table}

\subsection{Baryons}
In this section, we calculate the Schottky peaks of baryon.  First, we examine the light baryons, and next we consider the heavy baryons.  Finally we discuss the flavor dependence of the baryon energy scales from these results.

\subsubsection{Light baryons}
\begin{figure}
\includegraphics[width=7.0cm]{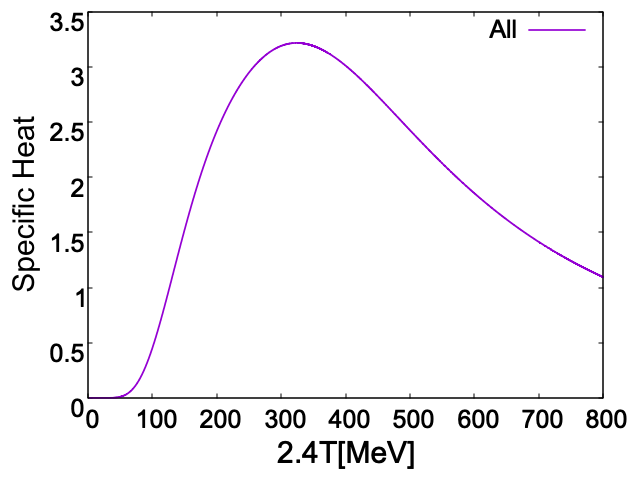}
\caption{Specific heat of light baryon \label{fig:lightbaryon}}
\end{figure}

We calculate the specific heat using all of the light baryons listed in Table~\ref{tab:lightnonsb} and \ref{tab:lightstrangeb}. In Fig.~\ref{fig:lightbaryon} we show the specific heats calculated with all the light baryons and find only one peak at 324~MeV. In this peak, several excitation energies are expected to be contributed because the hyperfine splitting can be larger than heavy quark systems according to Eq.~(\ref{eq:HF}).

\begin{figure}
\includegraphics[width=7.0cm]{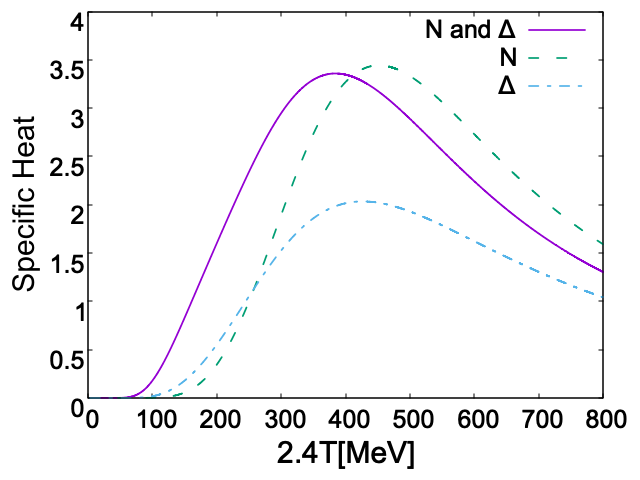}
\caption{Specific heats of the light unflavored baryons. The solid line stands for the specific heat calculated with all the unflavored baryons, while the dashed and dashed-dotted lines show those of the nucleons and the $\Delta$ baryons, respectively.  \label{fig:lightnosbaryon}}
\end{figure}

To resolve the compound excitations, we decompose the light baryons in terms of their strangeness. In Fig.~\ref{fig:lightnosbaryon}, we show the specific heat for the $N$ and $\Delta$ baryons in the solid line. Only one peak is seen at 384~MeV. Further, we decompose the spectrum into the $N$ and $\Delta$ sectors in order to resolve possible spin-spin interactions seen in the ground states of $N$ and $\Delta$. The decomposed specific heats are shown in the dashed line for $N$ and the dashed-dotted line for $\Delta$ of Fig.~\ref{fig:lightnosbaryon}. We find that the corresponding Schottky peaks are 452~MeV and 427~MeV, respectively. These values are larger than the peak position of the solid line where the $N$ and $\Delta$ are included into the calculation collectively. This is because the mass difference of $N(940)$ and $\Delta(1232)$, which is around 270~MeV, is also seen in the Schottky peak of the $N$-$\Delta$ spectrum. Although the $N(940)$-$\Delta(1232)$ mass difference stems from the hyperfine splitting, one cannot resolve all the effects of the hyperfine splitting in the decomposed specific heats.

\begin{figure}[htbp]
\begin{minipage}[b]{\linewidth}
\centering
\includegraphics[width=7.4cm]{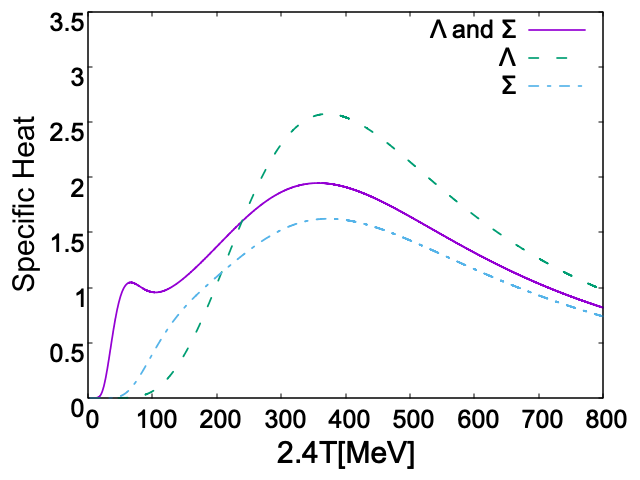}
\subcaption{\label{fig:LS}}
\end{minipage}\\
\begin{minipage}[b]{\linewidth}
\centering
\includegraphics[width=7.4cm]{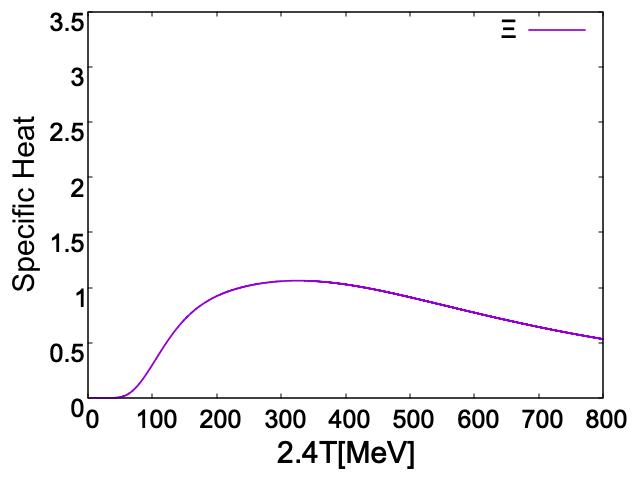}
\subcaption{\label{fig:xi}}
\end{minipage} \\
\caption{Specific heats of the hyperons. (a) Specific heats of the hyperons with strangeness $S=-1$. The solid line shows the specific heats for the $\Lambda$ and $\Sigma$ baryons, while the dashed and dashed-dotted lines stand for those for the $\Lambda$ and $\Sigma$ baryons, respectively. (b) Specific heat of the hyperons with strangeness $S=-2$. }
\label{fig:lightsbaryon}
\end{figure}

Next we consider hyperons with $S=-1$. The solid line in Fig.~\ref{fig:LS} shows the specific heat for the $\Lambda$ and $\Sigma$ spectrum and we find two peaks at 66~MeV and 358~MeV. The lower peak is considered to be the mass difference of $\Lambda(1115)$ and $\Sigma(1190)$, because this peak disappears when we consider the specific heat separately for $\Lambda$ and $\Sigma$ after performing the isospin decomposition. The separated specific heats for $\Lambda$ and $\Sigma$ are plotted as the dashed and dashed-dotted lines in Fig.~\ref{fig:LS} respectively. We find that they have only one Schottky peak located at 372~MeV and 373~MeV, respectively.  Since the specific heat of the $\Sigma$ system has a shoulder on the low temperature, it can be expected that an energy scale other than orbital excitation appears. This may be considered to be spin-spin splitting between $\Sigma(1190)$ and $\Sigma(1385)$.  But, we cannot separate this effect by the model independence way.

Finally, we will analyze the $\Xi$ system. Unfortunately there are only several states confirmed experimentally. The specific heat is shown in Fig.~\ref{fig:xi}, and the excitation energy scale is found to be 326~MeV. Only one peak appears, which is considered to be the average energy scale of hyperfine structure and orbital excitation like other baryons. We cannot separate the effect of the ground state hyperfine structure by a method with model independence.

\subsubsection{Heavy baryons}
\begin{figure}[htbp]
\begin{minipage}[b]{\linewidth}
\centering
\includegraphics[width=7.4cm]{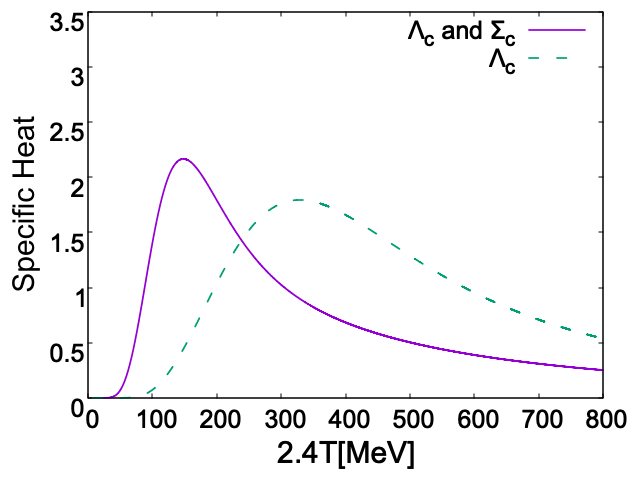}
\subcaption{\label{fig:LSc}}
\end{minipage}\\
\begin{minipage}[b]{\linewidth}
\centering
\includegraphics[width=7.4cm]{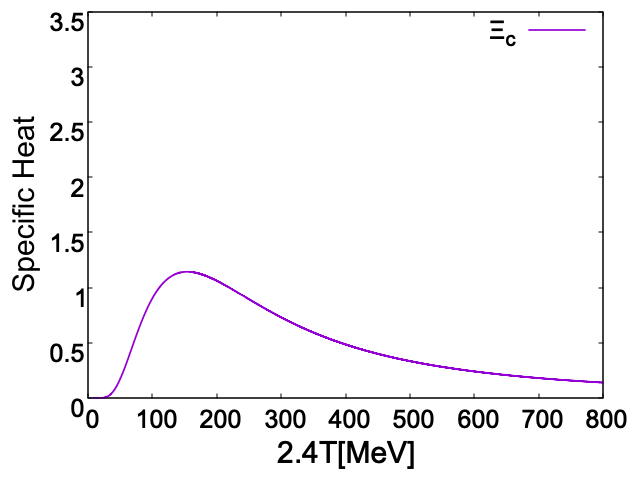}
\subcaption{\label{fig:Xic}}
\end{minipage} \\ 
\begin{minipage}[b]{\linewidth}
\centering
\includegraphics[width=7.4cm]{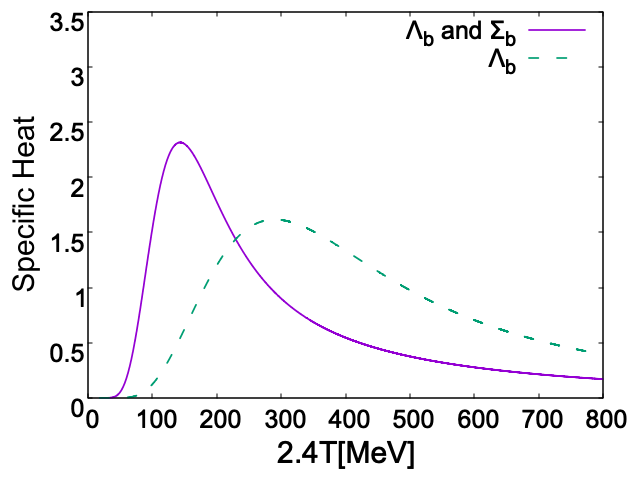}
\subcaption{\label{fig:LSb}}
\end{minipage}\\
\caption{Specific heats of the heavy baryons. (a) Specific heats of the charmed baryons with strangeness $S=0$. The solid line shows the specific heats for the $\Lambda_c$ and $\Sigma_c$ baryons, while the dashed line stands for that for the $\Lambda_c$ baryons only. (b) Specific heat of the $\Xi_c$ baryon. (c) Specific heats of the bottomed baryons with strangeness $S=0$. The solid line shows the specific heats for the $\Lambda_b$ and $\Sigma_b$ baryons, while the dashed line stands for that for the $\Lambda_b$ baryons only. }
\label{fig:heavybaryon}
\end{figure}

In this section, we investigate the Schottky peaks for heavy baryons. In Fig.~\ref{fig:LSc} we show the specific heat for the $\Lambda_c$ and $\Sigma_c$ baryons in the solid line. Here there are only one peak in the $\Lambda_c$-$\Sigma_c$ spectrum at 148~MeV. This peak may be contributed mainly by the mass difference between the ground states of $\Lambda_c$ and $\Sigma_c$ together with their internal excitations. In order to see the internal excitations in the $\Lambda_c$ baryon, we show separately the specific heat of the $\Lambda_c$ baryon in the dashed line of Fig.~\ref{fig:LSc} and find a Schottky peak at 329~MeV. For $\Sigma_c$, we do not have enough states to investigate the internal excitation energy. 

Figure~\ref{fig:Xic} shows the results of the calculation of the specific heats for the $\Xi_c$ system, which contains one charm quark and one strange quark. The energy scale is found to be 154~MeV. This peak may be considered to be a mixture of orbital excitation and hyperfine splitting as with other baryons.

Finally, the same analysis is performed for the $\Lambda_b$-$\Sigma_b$ system. The specific heat are shown in Fig.~\ref{fig:LSb} as the solid line and we find the energy scale to be 143~MeV. Again this may correspond to the mass difference between the ground states of $\Lambda_b$ and $\Sigma_b$. We also show the specific heat calculated with only the $\Lambda_b$ states in Fig.~\ref{fig:LSb} as the dashed line. This plot shows that the energy scale is 290~MeV.

\subsubsection{Discussion for the baryon system\label{sec:flavor}}

Table~\ref{tab:sumbaryon} summarizes the baryon excitation energies extracted by the Schottoky anomaly. As seen above, it is hard to separate out the hyperfine splitting in the baryon excitations in a model independent way. In order to investigate the flavor dependence of the energy scale of the baryons systematically, we compare the energy scales of $N$, $\Lambda$, $\Lambda_c$ and $\Lambda_b$. The $\Lambda$, $\Lambda_c$ and $\Lambda_b$ baryons have isospin $I=0$. This means that the $ud$ system in the ground states of $\Lambda$s  has the spin zero and the spin-spin force may not contribute to the mass spectrum. The ratios of the excitation energy to the nucleon one is found to be 1.00, 0.82, 0.73, 0.64 for $N$, $\Lambda$, $\Lambda_c$ and $\Lambda_b$, respectively. 

\begin{table}
\caption{Excitation energy scales of the baryons obtained by Schottky anomaly in units of MeV. Symbol $S$, $C$ and $B^\prime$ stand for strangeness, charm and bottom quantum numbers, respectively.\label{tab:sumbaryon}}
\begin{ruledtabular}
\begin{tabular}{llll}
Light baryons      & $S=0$      & $S=-1$      & $S=-2$       \\
All                & 384        & 66, 358     & 326          \\
$I=\frac{1}{2}, 0$ & 452        & 372         & -            \\
$I=\frac{3}{2}, 1$ & 427        & 373         & -            \\ \hline
Heavy baryons      & $C=1, S=0$ & $C=1, S=-1$ & $B'=-1, S=0$ \\
All                & 148        & 154         & 143          \\
$I=0$              & 329        & -           & 290         
\end{tabular}
\end{ruledtabular}
\end{table}

In order to extract the systematics of the excitation energies among the $N$, $\Lambda$, $\Lambda_c$ and $\Lambda_b$ baryons, we investigate the scaling law of the excitation energy. We assume that the excitation energy can be described by power of a mass scale of the system:
    \begin{equation}
       \Delta E(M) = \frac{a}{(M/\Lambda)^b}   \label{eq:deltaE}
    \end{equation}
where $a$ and $b$ are constant parameters, $M$ is a mass of interest and $\Lambda$ is a normalization of the mass scale. Here we set $\Lambda = 1$ MeV. We try several models to investigate the systematics of the excitation energy. 
  
First, we consider the excitation energy to be a function of the ground state mass $M_0$ for each system. We fit the excitation energies by Eq.~\eqref{eq:deltaE} with $M=M_0$ and find out the values of $a$ and $b$. The ground state masses used in the fitting are shown in Table~\ref{tab:flavordep}. The fitted values are $a = 2034$ MeV and $b=0.23$. This implies that $\Delta E \propto M_0^{-1/4}$. 
  
Next, we consider $\Delta E$ to be a function of a constituent quark mass $M_q$. These baryons have the up and down quarks in common and one of the quarks are different among them. Therefor, these baryons can be specified by $u$ or $d$, $s$, $c$ and $b$ quarks, respectively. The excitation energy can be regarded as a function of these constituent quarks. We fit the excitation energy by Eq.~\eqref{eq:deltaE} with $M=M_0$. In Table~\ref{tab:flavordep}, we show the values of the constituent quark masses used in the fitting and the fitted values of $a$ and $b$. The fitted value of $b$ tells us that $\Delta E \propto M_q^{-1/6}$.
  
\begin{figure}[htbp]
\begin{minipage}[b]{\linewidth}
\centering
\includegraphics[width=7.4cm]{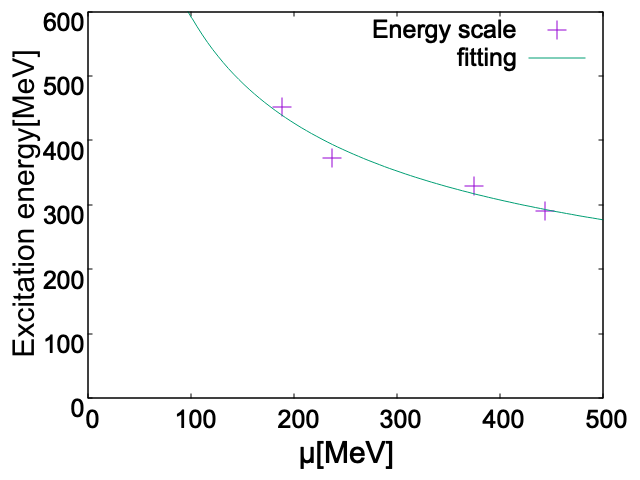}
\subcaption{\label{fig:e-rmass}}
\end{minipage}\\
\begin{minipage}[b]{\linewidth}
\centering
\includegraphics[width=7.4cm]{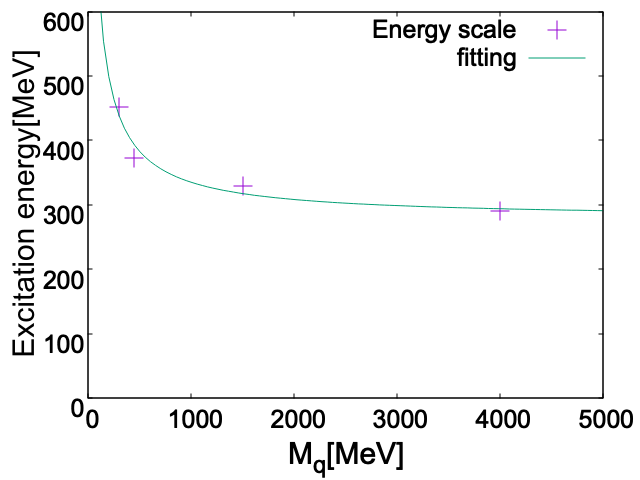}
\subcaption{\label{fig:dmass}}
\end{minipage}
\caption{(a) Excitation energy scales of baryon as a function of the reduced mass of the $ud$ diquark and the constituent quark. The points are the excitation energy obtained by the Schottky anomaly, while the line denotes the calculation using Eq. (\ref{eq:deltaE}) with the fitted values a = 5238 MeV and b = 0.47. The diquark mass is assumed to 500 MeV. The constituent quark masses used in the plot are shown in Table \ref{tab:flavordep}. (b) Excitation energy scales of baryon as a function of the constituent quark mass $M_q$. The line is calculated using Eq. (\ref{eq:HOdiquark}) with the fitted values  $k = 890$ MeV/fm$^2$ and $M_{dq} = 445$ MeV.}
\label{fig:baryonex}
\end{figure}

Finally, we take a diquark picture for these baryons. We regard the $u$ and $d$ quarks in the baryon to form a diquark and these baryon are considered to be a two-body system of the diquark and a quark. In this picture the excitation energy can be a function of the reduced mass $\mu$ of the diquark and the quark. Here we assume the constituent diquark mass as $M_{dq}=500$~MeV and the constituent quark masses as shown in Table \ref{tab:flavordep}. In Fig.~\ref{fig:e-rmass} we show the excitation energy as a function of the reduced mass. Fitting the excitation energy by Eq.~\eqref{eq:deltaE} with $M=\mu$, we obtain $a=5238$ MeV and $b=0.47$. This implies that $\Delta E \propto \mu^{-1/2}$. Because this dependence is same as a simple harmonic oscillator, the system can be regarded as a $ud$ diquark and quark two-body system with a harmonic oscillator potential. From the fitted value $a$ the string constant of the oscillator is found to be $k=707$~MeV/fm$^2$. 
  
Now assuming that these baryons are composed of a $ud$ diquark and a quark with a harmonic oscillator potential, we try to extract the string constant $k$ and the $ud$ diquark mass $M_{dq}$ from the excited energies. For the harmonic oscillator system the excitation energy is described as 
  \begin{equation}
    \Delta E = \sqrt{\frac k \mu}, \qquad {\rm with} \quad
      \mu = \frac{M_{dq}M_q}{M_{dq} + M_q}.
      \label{eq:HOdiquark}
  \end{equation}   
By fitting the excitation energies with Eq.~\eqref{eq:HOdiquark}, we find that $k=890$~MeV/fm$^2$ and $M_{dq} = 445$~MeV.  In Fig~\ref{fig:dmass} we show the excitation energies of the baryons extracted by the Schottky anomaly in points as a function of the constituent quark mass $M_q$ together with that obtained using the fitted values of $k$ and $M_{dq}$ in the solid line. 

\begin{table}
\caption{Parameters used for discussion in Sec. III.C.3. $\Delta E$ is the excitation energy scale obtained by the Schottky anomaly. $M_0$ is the grand state mass of each baryon. $M_q$ is an assumed values of the constituent quark mass of each baryon. $\mu$ is the reduced mass of the $ud$ diquark and the constituent quark mass, where the diquark mass is assumed to be $M_{dq}=500$ MeV.  \label{tab:flavordep}}
\begin{ruledtabular}
\begin{tabular}{lllll}
              & $\Delta E${[}MeV{]} & $M_{0}${[}MeV{]} & $M_{q}${[}MeV{]} & $\mu${[}MeV{]} \\
$N$           & 452                 & 940              & 300              & 188            \\
$\Lambda$     & 372                 & 1116             & 450              & 237            \\
$\Lambda_{c}$ & 329                 & 2286             & 1500             & 375            \\
$\Lambda_{b}$ & 290                 & 5620             & 4000             & 444            \\ \hline
$a${[}MeV{]}  & -                   & 2034             & 1073             & 5238           \\
$b$           & -                   & 0.2306           & 0.160            & 0.473         
\end{tabular}
\end{ruledtabular}
\end{table}

\subsubsection{Positive and negative parity baryons}

The $N(1440)$ nucleon resonance is the first excited state of nucleon with $J^{P} = \frac{1}{2}^{+}$.  This state is called Roper resonance \cite{roper0} and the mass cannot be reproduced by simple quark model \cite{roper1}.  The property of the Roper resonance is one of the important subject in hadron physics.  It was pointed out that similar lower excited states with positive parity  are also found for hyperons \cite{roper2}, and recently their heavy baryon candidates have been proposed in Ref.\ \cite{roper3}. It is very interesting that these Roper like excited states are located about 500 MeV higher than their ground state.  Here we separately calculate the excitation energies for the positive and negative party baryons. 

\begin{figure}[htbp]
\begin{minipage}[b]{\linewidth}
\centering
\includegraphics[width=7.4cm]{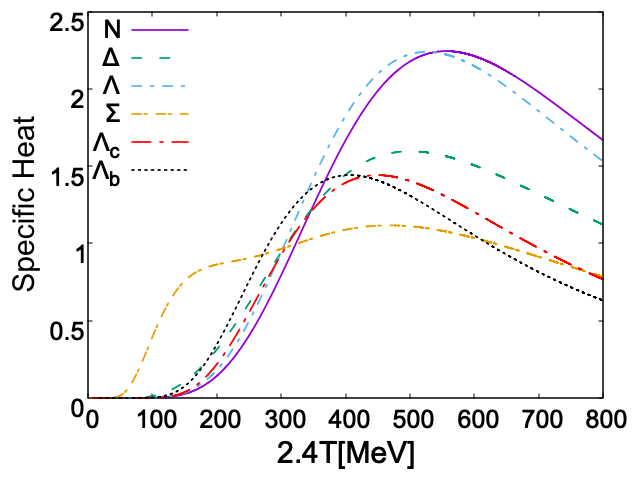}
\subcaption{\label{fig:posi}}
\end{minipage}\\
\begin{minipage}[b]{\linewidth}
\centering
\includegraphics[width=7.4cm]{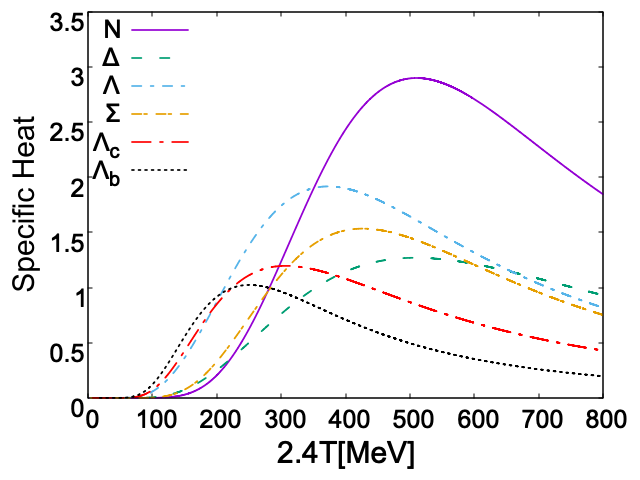}
\subcaption{\label{fig:nega}}
\end{minipage}
\caption{(a) Specific heat of the positive parity baryons.  The solid line, short dashed, short dashed-dotted, long dashed, long dashed-dotted and dotted lines show specific heat of the Nucleon, $\Delta$, $\Lambda$, $\Sigma$, $\Lambda_{c}$, and $\Lambda_{b}$ baryons, respectively .  (b)Same as (a) but for negative parity baryons.}
\label{fig:parity}
\end{figure}

For the $N$, $\Delta$, $\Lambda$, $\Sigma$, $\Lambda_{c}$ and $\Lambda_{b}$ baryons, we calculate the excitation energy scale of the positive and negative parity states using their specific heats.  Figure~\ref{fig:posi} shows the specific heats of the positive parity states.  We read the excitation energies from the figure and we summarized the excitation energy scales in Table~\ref{tab:parity}.  The obtained excited energies are around 500~MeV less dependently on their flavor. In order to see the excitation energy to negative parity states, we calculate the specific heat with the negative parity states together with the ground state. Figure~\ref{fig:nega} shows the specific heats and Table~\ref{tab:parity} summarizes the excitation energies extracted from the specific heats. From the table we find strong flavor dependence fo the excited energies to the negative parity states; the excitation energy gets smaller for the heavier baryons. This is consistent with what we have found in Sec.~\ref{sec:flavor}. Thus, we say that for baryons we have flavor dependence on the orbital excitation energies. 

\begin{table}
\caption{Excitation energy scales of the positive and negative parity baryons obtained by Schottky anomaly in units of MeV.\label{tab:parity}}
\begin{ruledtabular}
\begin{tabular}{lllllll}
                & $N$ & $\Delta$ & $\Lambda$ & $\Sigma$ & $\Lambda_{c}$ & $\Lambda_{b}$ \\
Positive Parity & 557 & 501      & 527       & 468      & 450           & 407           \\
Negative Parity & 510 & 509      & 373       & 427      & 307           & 251          
\end{tabular}
\end{ruledtabular}
\end{table}

\section{Summary}
The specific heats of the hadronic systems have been calculated using the hadron masses observed in experiment. We use Schottky anomaly to extract the typical excitation energy scales of a hadronic system $\Delta E$. The extraction is done by reading the temperature $T_{\rm{peak}}$ where the specific heat has a peak and by assuming $\Delta E \simeq 2.4T_{\rm{peak}}$, which is a general relation in the two-state system. 

We have obtained excitation energy scales for hadronic systems. Among the excitation energy scales, we have pinned down that of the hyperfine structure in the heavy meson systems and the excitation energy coming from the global structure of the confinement potential which is seen commonly in all of the hadron spectra. Thus, the latter scale may stem from orbital excitation. In the meson system, the energy scale of the orbital excitation is found to be 400 to 440~MeV insensitively to flavor dependence. On the other hand, in the baryon systems, we find flavor dependence of the energy scale of orbital excitation. The excitation energies for heavier systems are smaller than those for lighter systems. We have found by comparing the obtained excitation energies of $N$, $\Lambda$, $\Lambda_c$ and $\Lambda_b$ baryons, which have similar isospin structures, that the excitation energy for these baryons can be described as $\Delta E \propto \mu^{-1/2}$, where $\mu$ is the reduced mass of the consistent quark and a $ud$ diquark. This implies that these baryons may be described by a constituent quark and a $ud$ diquark with a harmonic oscillator potential.  We have also examined separately the excitation energies to the positive and negative parity excited states from the ground state for baryons. We have found that the positive parity excitation energy is rather flavor independent as pointed out in Ref.~\cite{roper3}, while the negative parity excitation has flavor dependence that the heavier baryons have smaller excitation energies than the lighter baryons.  As a future prospect, it would be interesting to confirm these conclusions if one could observe further excited states, such as open heavy mesons, $\Xi$, $\Xi_c$ and $\Xi_b$ baryons, in which lower excited states are not sufficiently observed.

\begin{acknowledgments}
The authors would like to thank Prof. Hosaka for his suggestion to calculate the excitation energies separately for positive and negative parity baryons.  The work of DJ was partially supported by the Grant-in-Aid for Scientific Research (Nos.\ JP17K05449 and 21K03530) from JSPS. 
\end{acknowledgments}

\appendix

\section{\label{sec:list}Tables of states used for calculation of specific heat}
In this section we list the hadron states used to calculate the specific heat.  We use the value in the bracket of the name of the state for its mass.

\begin{table*}
\caption{Charmonium states used in the calculation. In the left (right) line, the charmonia in which two quarks form spin 0 (1) are listed. (See the text for the details.)  \label{tab:ccbar}}
\begin{ruledtabular}
\begin{tabular}{llll}
$\eta_{c}, h_{c}$ & $J^{PC}$ & $\Psi, \chi_{c}$  &  $J^{PC}$   \\ \hline 
$\eta_{c}$(2984) & $0^{-+}$ & $J/\Psi$(3097) & $1^{--}$  \\ 
$h_{c}$(3525) & $1^{+-}$ & $\chi_{c0}$(3415) & $0^{++}$  \\ 
$\eta_{c}$(3638) & $0^{-+}$ & $\chi_{c1}$(3511) & $1^{++}$  \\ 
&& $\chi_{c2}$(3556) & $2^{++}$  \\ 
&& $\Psi$(3686) & $1^{--}$  \\ 
&& $\Psi$(3770) & $1^{--}$  \\ 
&& $\Psi_{2}$(3823) & $2^{--}$  \\
&& $\Psi_{3}$(3842) & $3^{--}$  \\
&& $\chi_{c1}$(3872) & $1^{++}$  \\
&& $\chi_{c2}$(3930) & $2^{++}$ \\
&& $\Psi$(4040) & $1^{--}$  \\
&& $\chi_{c1}$(4140) & $1^{++}$  \\
&& $\Psi$(4160) & $1^{--}$  \\
&& $\Psi$(4230) & $1^{--}$  \\
&& $\chi_{c1}$(4274) & $1^{++}$  \\
&& $\Psi$(4360) & $1^{--}$ \\ 
&& $\Psi$(4415) & $1^{--}$  \\ 
&& $\Psi$(4660) & $1^{--}$  \\ 
\end{tabular}
\end{ruledtabular}
\end{table*}

\begin{table*}
\caption{Bottomonium states used in the calculation. In the left (right) line, the bottomonia in which two quarks form spin 0 (1) are listed. (See the text for the details.) \label{tab:bbbar}}
\begin{ruledtabular}
\begin{tabular}{llll}
$\eta_{b}, h_{b}$  & $J^{PC}$ & $\Psi, \chi_{b}$  & $J^{PC}$   \\ \hline 
$\eta_{b}$(9399)  & $0^{-+}$ & $\Upsilon$(9460) & $1^{--}$  \\ 
$h_{b}$(9899)  & $1^{+-}$ & $\chi_{b0}$(9859) & $0^{++}$  \\ 
&& $\chi_{b1}$(9893) & $1^{++}$  \\ 
&& $\chi_{b2}$(9912) & $2^{++}$  \\ 
&& $\Upsilon$(10023) & $1^{--}$  \\ 
&& $\Upsilon_{2}$(10164) & $2{--}$  \\ 
&& $\chi_{b0}$(10233) & $0^{++}$  \\ 
&& $\chi_{b1}$(10255) & $1^{++}$  \\ 
&& $\chi_{b2}$(10269) & $2^{++}$  \\ 
&& $\Upsilon$(10355) & $1^{--}$  \\ 
&& $\chi_{b1}$(10513) & $1^{++}$  \\ 
&& $\chi_{b2}$(10524) & $1^{++}$  \\ 
&& $\Upsilon$(10579) & $1^{--}$  \\ 
&& $\Upsilon$(10860) & $1^{--}$  \\ 
&& $\Upsilon$(11020) & $1^{--}$  \\ 
\end{tabular}
\end{ruledtabular}
\end{table*}

\begin{table*}
\caption{Open chram states used in the calculation. \label{tab:openc}}
\begin{ruledtabular}
\begin{tabular}{llll}
$D, D^{*}$        & $J^{P}$ & $D_{s},D_{s}^{*} $ & $J^{P}$ \\ \hline
$D(1870)$         & $0^{-}$ & $D_{s}(1968)$      & $0^{-}$ \\
$D(2420)$         & $1^{+}$ & $D_{s1}(2460)$     & $1^{+}$ \\
                  &         & $D_{s1}(2536)$     & $1^{+}$ \\
$D^{*}(2007)$     & $1^{-}$ & $D_{s}^{*}(2112)$  & $1^{-}$ \\
$D^{*}_{0}(2300)$ & $0^{+}$ & $D_{s0}^{*}(2317)$ & $0^{+}$ \\
$D^{*}_{2}(2460)$ & $2^{+}$ & $D_{s2}^{*}(2573)$ & $2^{+}$ \\
                  &         & $D_{s1}^{*}(2700)$ & $1^{-}$
\end{tabular}
\end{ruledtabular}
\end{table*}

\begin{table*}
\caption{Open bottom states used in the calculation. \label{tab:openb}}
\begin{ruledtabular}
\begin{tabular}{llll}
$B, B^{*}$        & $J^{P}$ & $B_{s},B_{s}^{*} $ & $J^{P}$ \\ \hline
$B(5279)$         & $0^{-}$ & $B_{s}(5367)$      & $0^{-}$ \\
$B(5721)$         & $1^{+}$ & $B_{s1}(5830)$     & $1^{+}$ \\
$B^{*}(5325)$     & $1^{-}$ & $B_{s}^{*}(5425)$  & $1^{-}$ \\
$B^{*}_{2}(5747)$ & $2^{+}$ & $B_{s2}^{*}(5840)$ & $2^{+}$ \\
\end{tabular}
\end{ruledtabular}
\end{table*}

\begin{table*}
\caption{Light meson states used in the calculation.\label{tab:lightmeson}}
\begin{ruledtabular}
\begin{tabular}{llllll}
\multicolumn{2}{l}{$I=1$}    & \multicolumn{2}{l}{$I=0$}                     & \multicolumn{2}{l}{$I=\frac{1}{2}$} \\
$\pi, \rho, a, b$ & $J^{PC}$ & $\eta,\eta',\omega,\phi,h,h',f,f'$ & $J^{PC}$ & $K,K^{*}$             & $J^{P}$     \\ \hline
$\pi(135)$        & $0^{-+}$ & $f_{0}(500)$                       & $0^{++}$ & $K(498)$              & $0^{-}$     \\ 
$\rho(770)$       & $1^{--}$ & $\eta(547)$                        & $0^{-+}$ & $K_{0}^{*}(700)$      & $0^{+}$     \\
$a_{0}(980)$      & $0^{++}$ & $\omega(782)$                      & $1^{--}$ & $K^{*}(892)$          & $1^{-}$     \\
$b_{1}(1235)$     & $1^{+-}$ & $\eta'(958)$                       & $0^{-+}$ & $K_{1}(1270)$         & $1^{+}$     \\
$a_{1}(1260)$     & $1^{++}$ & $f_{0}(980)$                       & $0^{++}$ & $K_{1}(1400)$         & $1^{+}$     \\
$\pi(1300)$       & $0^{-+}$ & $\phi(1020)$                       & $1^{--}$ & $K^{*}(1410)$         & $0^{+}$     \\
$a_{2}(1320)$     & $2^{++}$ & $h_{1}(1170)$                      & $1^{+-}$ & $K_{0}^{*}(1430)$     & $0^{+}$     \\
$\pi_{1}(1400)$   & $1^{-+}$ & $f_{2}(1270)$                      & $2^{++}$ & $K_{2}^{*}(1430)$     & $2^{+}$     \\
$a_{0}(1450)$     & $0^{++}$ & $f_{1}(1285)$                      & $1^{++}$ & $K^{*}(1680)$         & $1^{-}$     \\
$\rho(1450)$      & $1^{--}$ & $\eta(1295)$                       & $0^{-+}$ & $K_{2}(1770)$         & $2^{-}$     \\
$\pi_{1}(1600)$   & $1^{-+}$ & $f_{0}(1370)$                      & $0^{++}$ & $K_{3}^{*}(1780)$     & $3^{-}$     \\
$a_{1}(1640)$     & $1^{++}$ & $\eta(1405)$                       & $0^{-+}$ & $K_{2}(1820)$         & $2^{-}$     \\
$\pi_{2}(1670)$   & $2^{-+}$ & $h_{1}(1415)$                      & $1^{+-}$ & $K_{4}^{*}(2045)$     & $4^{+}$     \\
$\rho_{3}(1690)$  & $3^{--}$ & $f_{1}(1420)$                      & $1^{++}$ &                       &             \\
$\rho(1700)$      & $1^{--}$ & $\omega(1420)$                     & $1^{--}$ &                       &             \\
$a_{2}(1700)$     & $2^{++}$ & $\eta(1475)$                       & $0^{-+}$ &                       &             \\
$\pi(1800)$       & $0^{-+}$ & $f_{0}(1500)$                      & $0^{++}$ &                       &             \\
$\pi_{2}(1880)$   & $2^{-+}$ & $f_{2}'(1525)$                     & $2^{++}$ &                       &             \\
$a_{4}(1970)$     & $4^{++}$ & $\eta_{2}(1645)$                   & $2^{-+}$ &                       &             \\
                  &          & $\omega(1650)$                     & $1^{--}$ &                       &             \\
                  &          & $\omega_{3}(1670)$                 & $3^{--}$ &                       &             \\
                  &          & $\phi(1680)$                       & $1^{--}$ &                       &             \\
                  &          & $f_{0}(1710)$                      & $0^{++}$ &                       &             \\
                  &          & $\phi_{3}(1850)$                   & $1^{--}$ &                       &             \\
                  &          & $\eta_{2}(1870)$                   & $2^{-+}$ &                       &             \\
                  &          & $f_{2}(1950)$                      & $2^{++}$ &                       &             \\
                  &          & $f_{2}(2010)$                      & $2^{++}$ &                       &             \\
                  &          & $f_{4}(2050)$                      & $4^{++}$ &                       &             \\
                  &          & $\phi(2170)$                       & $1^{--}$ &                       &             \\
                  &          & $f_{2}(2300)$                      & $2^{++}$ &                       &             \\
                  &          & $f_{2}(2340)$                      & $2^{++}$ &                       &            
\end{tabular}
\end{ruledtabular}
\end{table*}

\begin{table*}
\caption{Light unflavored baryon states used in the calculation. \label{tab:lightnonsb}}
\begin{ruledtabular}
\begin{tabular}{llll}
$N$($I=1/2$) & $J^{P}$ & $\Delta$($I=3/2$) & $J^{P}$ \\ \hline 
$N(940)$ & ${\frac{1}{2}}^{+}$ & $\Delta$(1232) & $\frac{3}{2}^{+}$ \\ 
$N(1440)$ & ${\frac{1}{2}}^{+}$ & $\Delta$(1600) & $\frac{3}{2}^{+}$  \\ 
$N(1520)$ & ${\frac{3}{2}}^{-}$ & $\Delta$(1620) & $\frac{1}{2}^{-}$  \\ 
$N(1535)$ & ${\frac{1}{2}}^{-}$ & $\Delta$(1700) & $\frac{3}{2}^{-}$  \\ 
$N(1650)$ & ${\frac{1}{2}}^{-}$ & $\Delta$(1900) & $\frac{1}{2}^{-}$  \\ 
$N(1675)$ & ${\frac{5}{2}}^{-}$ & $\Delta$(1905) & $\frac{5}{2}^{+}$ \\ 
$N(1680)$ & ${\frac{5}{2}}^{+}$ & $\Delta$(1910) & $\frac{1}{2}^{+}$  \\ 
$N(1700)$ & ${\frac{3}{2}}^{-}$ & $\Delta$(1920) & $\frac{3}{2}^{+}$  \\ 
$N(1710)$ & ${\frac{1}{2}}^{+}$ & $\Delta$(1930) & $\frac{5}{2}^{-}$   \\ 
$N(1720)$ & ${\frac{3}{2}}^{+}$ & $\Delta$(1950) & $\frac{7}{2}^{+}$  \\ 
$N(1875)$ & ${\frac{3}{2}}^{-}$ & $\Delta$(2200) & $\frac{7}{2}^{-}$   \\ 
$N(1880)$ & ${\frac{1}{2}}^{+}$ & $\Delta$(2420) & $\frac{11}{2}^{+}$ \\ 
$N(1895)$ & ${\frac{1}{2}}^{-}$ & & \\ 
$N(1900)$ & ${\frac{3}{2}}^{+}$ & & \\ 
$N(2060)$ & ${\frac{5}{2}}^{-}$ & & \\ 
$N(2100)$ & ${\frac{1}{2}}^{+}$ & & \\ 
$N(2120)$ & ${\frac{3}{2}}^{-}$ & & \\ 
$N(2190)$ & ${\frac{7}{2}}^{-}$ & & \\ 
$N(2220)$ & ${\frac{9}{2}}^{+}$ & & \\ 
$N(2250)$ & ${\frac{9}{2}}^{-}$ & & \\ 
$N(2600)$ & ${\frac{11}{2}}^{-}$ & & \\ 
\end{tabular}
\end{ruledtabular}
\end{table*}

\begin{table*}
\caption{Light strange baryon states used in the calculation.  \label{tab:lightstrangeb}}
\begin{ruledtabular}
\begin{tabular}{llllll}
$\Lambda$($I=0$) & $J^{P}$ & $\Sigma$($I=1$) & $J^{P}$ & $\Xi$($I=1/2$), $\Omega$($I=0$) & $J^{P}$ \\ \hline
$\Lambda$(1115) & $\frac{1}{2}^{+}$ & $\Sigma$(1190) & $\frac{1}{2}^{+}$ & $\Xi$(1315) & $\frac{1}{2}^{+}$ \\ 
$\Lambda$(1405) & $\frac{1}{2}^{-}$ & $\Sigma$(1385) & $\frac{3}{2}^{+}$ & $\Xi$(1530) & $\frac{3}{2}^{+}$ \\ 
$\Lambda$(1520) & $\frac{3}{2}^{-}$ & $\Sigma$(1660) & $\frac{1}{2}^{+}$ & $\Xi$(1820) & $\frac{1}{2}^{-}$ \\ 
$\Lambda$(1600) & $\frac{1}{2}^{+}$ & $\Sigma$(1670) & $\frac{3}{2}^{-}$ & $\Xi$(2030) & $\frac{5}{2}^{?}$ \\
$\Lambda$(1670) & $\frac{1}{2}^{-}$ & $\Sigma$(1750) & $\frac{1}{2}^{-}$ & $\Omega$(1672) & $\frac{3}{2}^{+}$ \\
$\Lambda$(1690) & $\frac{3}{2}^{-}$ & $\Sigma$(1775) & $\frac{5}{2}^{-}$ && \\ 
$\Lambda$(1800) & $\frac{1}{2}^{-}$ & $\Sigma$(1910) & $\frac{3}{2}^{-}$ && \\ 
$\Lambda$(1810) & $\frac{1}{2}^{+}$ & $\Sigma$(1915) & $\frac{5}{2}^{+}$ && \\
$\Lambda$(1820) & $\frac{5}{2}^{+}$ & $\Sigma$(2030) & $\frac{7}{2}^{+}$ && \\
$\Lambda$(1830) & $\frac{5}{2}^{-}$ &&&& \\
$\Lambda$(1890) & $\frac{3}{2}^{+}$ &&&& \\
$\Lambda$(2100) & $\frac{7}{2}^{-}$ &&&& \\
$\Lambda$(2110) & $\frac{5}{2}^{+}$ &&&& \\
$\Lambda$(2350) & $\frac{9}{2}^{+}$ &&&& \\
\end{tabular}
\end{ruledtabular}
\end{table*}

\begin{table*}
\caption{Heavy baryon states used in the calculation. \label{tab:heavyb}}
\begin{ruledtabular}
\begin{tabular}{llllll}
$\Lambda_{c}$, $\Sigma_{c}$ & $J^{P}$ & $\Xi_{c}$ & $J^{P}$ & $\Lambda_{b}$, $\Sigma_{b}$ & $J^{P}$ \\ \hline
$\Lambda_{c}$(2286) & $\frac{1}{2}^{+}$ & $\Xi_{c}$(2470) & $\frac{1}{2}^{+}$ & $\Lambda_{b}$(5620) & $\frac{1}{2}^{+}$ \\
$\Lambda_{c}$(2595) & $\frac{1}{2}^{-}$ & $\Xi_{c}$(2790) & $\frac{1}{2}^{-}$ & $\Lambda_{b}$(5912) & $\frac{1}{2}^{-}$ \\
$\Lambda_{c}$(2625) & $\frac{3}{2}^{-}$ & $\Xi_{c}$(2815) & $\frac{3}{2}^{-}$ & $\Lambda_{b}$(5920) & $\frac{3}{2}^{-}$ \\
$\Lambda_{c}$(2860) & $\frac{3}{2}^{+}$ & & & $\Lambda_{b}$(6146) & $\frac{3}{2}^{+}$ \\
$\Lambda_{c}$(2880) & $\frac{5}{2}^{+}$ & & & $\Lambda_{b}$(6152) & $\frac{5}{2}^{+}$ \\
$\Lambda_{c}$(2940) & $\frac{3}{2}^{-}$ & & & & \\
$\Sigma_{c}$(2455) & $\frac{1}{2}^{+}$ & $\Xi_{c}'$(2580) & $\frac{1}{2}^{+}$ & $\Sigma_{b}$(5810) & $\frac{1}{2}^{+}$ \\
$\Sigma_{c}$(2520) & $\frac{3}{2}^{+}$ & $\Xi_{c}'$(2645) & $\frac{3}{2}^{+}$ & $\Sigma_{b}$(5830) & $\frac{3}{2}^{+}$ \\
\end{tabular}
\end{ruledtabular}
\end{table*}


\begin{thebibliography}{99}
\bibitem{1}
E.~Eichten, K.~Gottfried, T.~Kinoshita, J.~B.~Kogut, K.~D.~Lane and T.~M.~Yan,
  Phys.\ Rev.\ Lett.\  {\bf 34}, 369 (1975),
  Erratum: [Phys.\ Rev.\ Lett.\  {\bf 36}, 1276 (1976)]; 
\bibitem{2}
S.N. Mukherjee, R. Nag, S. Sanyal, T. Morii, J. Morishita, and M. Tsuge, Phys.Rept. {\bf 231}, 201 (1993).
\bibitem{3}
N.~Isgur, G.~Karl, Phys.\ Rev.\ D {\bf 18}, 4187 (1978); {\bf 20}, 1191 (1979); 
\bibitem{4}
S.~Capstick, W.~Roberts, Phys.\ Rev.\ D {\bf 49}, 4570 (1994).
\bibitem{5}
D.~Jido and M.~Sakashita, PTEP {\bf 2016}, 083D02 (2016); 
\bibitem{6}
K.~Kumakawa and D.~Jido, PTEP {\bf 2017}, 123D01 (2017); [arXiv:2107.11950 [hep-ph]].
\bibitem{7}
R.K.~Bhaduri and M.~Dey, Phys.\ Lett.\ B125, 513 (1983).
\bibitem{8}
R.K.~Bhaduri and W.~van Dijk, Nucl.\ Phys.\ A485 1, (1988).
\bibitem{schotano}
A. Biswas, M.V.N. Murthy, N.Sinta, Phys. Rev. D92, 114012 (2015).
\bibitem{pdg}
P.A. Zyla et al. (Particle Data Group), Prog. Theor. Exp. Phys. 2020, 083C01 (2020)
\bibitem{particlephys}
A.D. R\'{u}jula, H. Georgi, S.L. Glashow, Phys. Rev. D12, 147 (1975).
\bibitem{roper0}
L.D. Roper, Phys. Rev. Lett 12, 34 (1964).
\bibitem{roper1}
V. D. Burkert and C. D. Roberts, Rev. Mod. Phys. 91, 011003 (2019).
\bibitem{roper2}
M. Takayama, H. Toki, and A. Hosaka, Prog. Theor. Phys. 101, 1271 (1999).
\bibitem{roper3}
A.J. Arifi, H. Nagahiro, A. Hosaka, K. Tanida, Phys. Rev. D 101, 111502 (2020).
\end{thebibliography}
\bibliographystyle{plain}

\end{document}